\documentclass[10pt,a4paper]{article}
\usepackage{amssymb} 
\usepackage{amsmath}
\usepackage{graphicx}
\usepackage{rsfs}    

\newcommand{\vect}[1]{\mbox{\boldmath $ #1 $}}
\newcommand{\bra}{\langle}
\newcommand{\ket}{\rangle}

\newcommand{\tr}{\mbox{Tr}}

\begin{document}
\hfill
arXiv: cond-mat.soft/0805.2471v2
\vspace*{16mm}
\begin{center}
{\Large \bf
Characterization of geometric structures 
\vspace{2mm} \\
of biaxial nematic phases
}
\vspace{12mm}\\
Shogo Tanimura\footnote{{\tt e-mail: tanimura@i.kyoto-u.ac.jp}}
\hspace{4mm} and \hspace{4mm}
Tomonori Koda\footnote{{\tt e-mail: koda@yz.yamagata-u.ac.jp}}
\vspace{2mm}\\
${}^1 ${\it
Department of Applied Mathematics and Physics,
Graduate School of Informatics,
\\
Kyoto University, Kyoto 606-8501, Japan
}
\vspace{2mm}\\
${}^2 ${\it
Graduate School of Science and Engineering, 
\\
Yamagata University, Yonezawa 992-8510, Japan
}
\vspace{8mm} \\
submitted on 16 May 2008, revised on 22 May 2008
\vspace{8mm}\\
Abstract
\vspace{2mm}\\
\begin{minipage}[t]{130mm}
\baselineskip 5.0mm
The ordering matrix, which was originally introduced by de Gennes,
is a well-known mathematical device 
for describing orientational order of biaxial nematic liquid crystal.
In this paper we propose a new interpretation of the ordering matrix.
We slightly modify the definition of the ordering matrix 
and call it the geometric order parameter.
The geometric order parameter is a linear transformation
which transforms a tensorial quantity of an individual molecule
to a tensorial quantity observed at a macroscopic scale.
The degree of order is defined as the singular value of the geometric order parameter.
We introduce the anisotropy diagram, which is useful 
for classification and comparison of various tensorial quantities.
As indices for evaluating anisotropies of tensorial quantities,
we define the degree of anisotropy and the degree of biaxiality.
We prove that a simple diagrammatic relation holds
between a microscopic tensor and a macroscopic tensor.
We provide a prescription to formulate the Landau-de Gennes free energy
of a system whose constituent molecules have an arbitrary shape.
We apply our prescription to a system which consists of $ D_{2h} $-symmetric molecules.
\end{minipage}
\vspace*{20mm}

\begin{minipage}{140mm}
\baselineskip 5.0mm 
PACS codes: 61.30.Cz, 61.30.Gd, 83.80.Xz
\vspace{1mm} 
\end{minipage}

\end{center}
\newpage
\baselineskip 4.76mm

\section{Introduction}
A system which consists of asymmetric molecules can exhibit various noticeable phenomena.
In particular, biaxial nematic liquid crystals are interesting subjects of current research.
Biaxiality means that physical properties 
of an individual molecule or an ensemble of molecules 
are not invariant under any rotations.
On the other hand, uniaxiality means that 
properties of a molecule or an ensemble of molecules 
are invariant under rotations about a fixed axis.
Anisotropy is a general concept which implies either uniaxiality or biaxiality.
Although most of molecules are not uniaxially symmetric in a rigorous sense,
it is possible that some properties of molecules are effectively uniaxial.
In a simple nematic liquid crystal
uniaxial molecules are aligned to exhibit a uniaxial order at a macroscopic scale.
However, in a more complex system under some circumstance,
it may happen that uniaxial molecules exhibit a biaxial order,
or 
it may also happen that biaxial molecules exhibit a biaxial order.

Biaxial nematic phases have been intensively studied in liquid crystal physics.
Williams~\cite{Williams1969} noticed 
biaxial anisotropy in optical properties of nematic liquid crystals in a magnetic field.
Taylor, Fergason, and Arora~\cite{Taylor1970} found a biaxial smectic C phase.
Freiser~\cite{Freiser1970} began a theoretical study of phase structures of asymmetric molecules
and predicted the existence of a biaxial nematic phase.
Alben~\cite{Alben1973} calculated the Landau free energy of a system of biaxial molecules
and predicted the existence of a biaxial phase.
Straley~\cite{Straley1974}
introduced four order parameters, $ (S,T,U,V) $ in his notation,
to describe nematic order structures of an ensemble of molecules 
which have the point group $ D_{2h} $ as their symmetry.
de Gennes and Prost~\cite{de Gennes-Prost1974} introduced a set of 
generalized order parameters, which is called the ordering matrix, 
to describe nematic order structures of molecules which have an arbitrary shape.
Yu and Saupe~\cite{Yu_Saupe1980} observed a biaxial phase in an experiment
and obtained a phase diagram in the concentration-temperature coordinate.
There the biaxial phase appeared between two distinct uniaxial phases.
Boonbrahm and Saupe~\cite{Saupe1984} studied effects of temperature and magnetic field 
on a thin film of biaxial nematic liquid.
Allender, Lee, and Hafiz~\cite{Allender1984, Allender1985}
constructed the Landau-de Gennes free energy of $ D_{2h} $-symmetric molecules
up to the sixth order in the Straley variables $ (S,T,U,V) $.
Bunning, Crellin, and Faber~\cite{Bunning1986}
studied experimentally an effect of molecular biaxiality on bulk properties,
particularly on the magnetic anisotropy.
Gramsbergen, Longa, and de Jeu~\cite{Gramsbergen1986} wrote a review on
the Landau theory of the nematic-isotropic phase transition.
However, the effects of biaxiality of molecules 
were not sufficiently considered in their review.
Remler and Haymet~\cite{Remler1986} gave a complete formulation
for describing interactions between asymmetric molecules
and applied their formulation to the analysis of the Landau free energy.
Mulder~\cite{Mulder1986} formulated
a model of sphero-platelet molecules which interact by exclusive volume effect.
Since his model has the $ D_{2h} $ symmetry, 
the order is described by the four order parameters of Straley.
Solving the problem by the mean field approximation,
he showed that a transition between the isotropic phase and the biaxial phase can occur.

The discoveries of a thermotropic biaxial phase
in a system of bent molecules (banana-shaped or boomerang-shaped molecules)
by Madsen, Dingemans, Nakata, and Samulski~\cite{Madsen2004}
and by Acharya, Primak, and Kumar~\cite{Acharya2004}
renewed the interest in biaxial nematics~\cite{Luckhurst2004}.
In their experiments it was observed that
biaxial molecules exhibited biaxial orders 
without application of external fields nor boundary effects.
Their discoveries have been stimulating intensive researches 
in this field~\cite{Merkel2004, Bates2005, Longa2005, Luckhurst2005, Longa2007, Allnder2007}.
Merkel {\it et al.}~\cite{Merkel2004} 
measured biaxiality parameters by infrared absorbance measurements
and compared the observed data with a result of the Landau-de Gennes model.
Bates and Luckhurst~\cite{Bates2005}
studied the phase diagram of a liquid which consists of V-shaped molecules
by the Monte Carlo simulation.
They showed existence of biaxial phases in the diagrams whose coordinates
are temperature 
and various anisotropy parameters like the bending angle of the molecule.

However, for a theoretical analysis of biaxial nematic phases,
it seems that there is still a confusion in descriptions of anisotropies.
In other words, it is necessary to invent a more useful and comprehensive method 
for describing anisotropies of molecules and nematic phases.
Let us discuss issues which exist in the present method 
for describing anisotropies of general nematics.
For a liquid crystal which consists of uniaxial molecules,
a well-known device to characterize orientational order of nematic phase is 
the tensorial order parameter
\begin{equation}
	\vect{A} = 
	\Big\langle \vect{n} \otimes \vect{n} - \frac{1}{3} I \Big\rangle.
	\label{A}
\end{equation}
Here $ \vect{n} $ is a unit vector 
which is fixed along the axis of each molecule and viewed from a laboratory observer.
Since the molecules execute thermal motion, 
the direction of the vector $ \vect{n} $ fluctuates.
The brackets $ \bra \cdots \ket $ mean a statistical average. 
The components of the tensor $ \vect{A} $ are written as
\begin{equation}
	A_{ij} = 
	\Big\langle n_i \, n_j - \frac{1}{3} \, \delta_{ij} \Big\rangle,
	\qquad
	i,j=1,2,3.
\end{equation}
The eigenvectors and eigenvalues of $ \vect{A} $ indicate alignment of the molecules.
The matrix $ \vect{A} $ can be diagonalized and parameterized as
\begin{equation}
	\vect{A} = 
	\frac{1}{3}
	\begin{pmatrix}
	- \sigma + \tau & 0 & 0 \\
	0 & - \sigma - \tau & 0 \\
	0 & 0 & 2 \sigma 
	\end{pmatrix}.
\end{equation}
When $ \sigma = \tau = 0 $, the system is in an isotropic phase.
When $ \sigma \ne 0 $ and $ \tau = 0 $, the system is in a uniaxial nematic phase.
The value of $ \sigma $ is in the range $ - \frac{1}{2} \le \sigma \le 1 $.
When $ \tau \ne 0 $, the system is in a biaxial nematic phase.

If the molecule itself is biaxial, we may introduce another order parameter,
\begin{equation}
	\vect{B} =
	\Big\langle \vect{l} \otimes \vect{l} - \vect{m} \otimes \vect{m}
	\Big\rangle.
	\label{B}
\end{equation}
Here $ \vect{l}, \vect{m}, \vect{n} $ are mutually orthogonal unit vectors fixed on each molecule.
The quantity $ \vect{B} $ characterizes 
the biaxial anisotropy of the nematic phase.
In general, we may associate principal values $ \lambda_1, \lambda_2, \lambda_3 $
with the principal axes $ \vect{l}, \vect{m}, \vect{n} $ 
and define the order parameter
\begin{equation}
	\vect{C} =
	\Big\langle 
	\lambda_1 \vect{l} \otimes \vect{l} +
	\lambda_2 \vect{m} \otimes \vect{m} +
	\lambda_3 \vect{n} \otimes \vect{n} -
	\frac{1}{3} ( \lambda_1 + \lambda_2 + \lambda_3 ) I
	\Big\rangle.
	\label{C}
\end{equation}

These order parameters, $ \vect{A} $, $ \vect{B} $, and $ \vect{C} $, are useful
for characterizing orientational order structures of nematic phases.
But there are several difficulties 
in their application to molecules which have an arbitrary shape.
First, there is no a priori reason to choose the molecular axes 
$ \vect{l}, \vect{m}, \vect{n} $
for an asymmetric molecule.
If the molecule is rectangular, choice of the axes is rather obvious.
However, for a molecule which has no symmetry,
choice of the axes is not unique.
There are various candidates for the molecular axes;
we may take the principal axes of 
the inertia tensor,
the dielectric susceptibility tensor,
the electric quadrupole tensor, or
the magnetic susceptibility tensor of the molecule.
In general, the axes defined by them do not coincide.
Thus, there is no unique definition of the molecular axes.
Second, distinction between the uniaxiality and the biaxiality becomes ambiguous
since the eigenvalues of $ \vect{A} $ and $ \vect{B} $ depend on the choice of the molecular axes.
Furthermore, there is no reason to choose a unique set of the principal values
$ \lambda_1, \lambda_2, \lambda_3 $ in the definition of the tensor $ \vect{C} $.
Third, the relation 
between the anisotropy of a molecule and the anisotropy of a macroscopic phase 
is vague in this kind of analysis.
It can happen that uniaxial molecules exhibit a biaxial phase.
It is also possible that biaxial molecules exhibit a uniaxial phase.
Thus a systematic method 
to compare the molecular anisotropy and the macroscopic anisotropy
is desirable.

de Gennes~\cite{de Gennes-Prost1974} introduced the ordering matrix
\begin{equation}
	S_{abij} 
	=
	\frac{3}{2}
	\left\langle
	R_{ai} \, R_{bj} - \frac{1}{3} \delta_{ab} \, \delta_{ij}
	\right\rangle,
	\qquad a,b,i,j = 1,2,3
	\label{de Gennes}
\end{equation}
for characterizing alignment of molecules in a nematic phase.
Here $ R_{ai} = \vect{L}_a \cdot \vect{M}_i $
is an inner product of the laboratory orthogonal frame 
$ ( \vect{L}_1, \vect{L}_2, \vect{L}_3 ) $
with the molecular orthogonal frame
$ ( \vect{M}_1, \vect{M}_2, \vect{M}_3 ) $.
The symmetrized tensor
\begin{equation}
	G_{abij} 
	=
	\left\langle
	\frac{1}{2} ( R_{ai} \, R_{bj} + R_{bi} \, R_{aj} )
	- \frac{1}{3} \delta_{ab} \, \delta_{ij}
	\right\rangle
	\label{ordering matrix}
\end{equation}
is more useful and meaningful as will be shown in this paper.
Although the ordering matrix is applicable to molecules of an arbitrary shape,
it is still difficult to read out geometrical and physical implications
from the ordering matrix.

In this paper we introduce a new approach for characterization and analysis
of anisotropies of a molecule and a bulk phase.
However, here we describe the outline of this paper.
In our discussion, the adjective {\it microscopic} means 
intrinsic properties or quantities which an individual molecule possesses.
On the other hand, {\it macroscopic} means 
average properties or quantities
observed in an ensemble of a large number of molecules.
If each molecule has a tensorial quantity $ t_{ij} $
and if the molecule changes its direction,
the tensor is transformed to 
$ \tilde{t}_{ab} = \sum_{i,j} R_{ai} \, R_{bj} \, t_{ij} $
by a rotation matrix $ R_{ai} $.
We assume that $ t_{ij} $ is a traceless symmetric tensor.
The quantity observable at a macroscopic scale is a statistical average
$ \bra \tilde{t}_{ab} \ket 
= \sum_{i,j} \bra R_{ai} \, R_{bj} \ket \, t_{ij} 
= \sum_{i,j} G_{abij} \, t_{ij} $.
This is an equation defining the {\it geometric order parameter} $ G_{abij} $.
Thus, the geometric order parameter can be regarded as a bridge which relates 
the microscopic quantity $ t_{ij} $ 
to the macroscopic quantity $ \bra \tilde{t}_{ab} \ket $.
Since the geometric order parameter $ G = ( G_{abij} ) $ is a linear transformation
$ \vect{t} \mapsto G \vect{t} $,
its property is completely analyzed by the method of singular value decomposition.
In Sect.~\ref{sect2} we will introduce the geometric order parameter
and discuss its properties.

After understanding the geometric order parameter,
the remaining task is to characterize 
anisotropies implied by the individual tensors,
$ t_{ij} $ and $ \bra \tilde{t}_{ab} \ket $.
To visualize the anisotropic property of a tensor
we introduce an {\it anisotropy diagram},
in which each tensor is represented as a point in a plane.
Then we define the {\it degree of anisotropy} $ \alpha ( \vect{t} ) $
and the {\it degree of biaxiality} $ \beta ( \vect{t} ) $
of the tensor $ \vect{t} = ( t_{ij} ) $.
Sect.~\ref{sect3} is an introductory discussion for providing the indices of anisotropy
and Sect.~\ref{sect4} is an explanation of the anisotropy diagram.

Furthermore, the geometric order parameter enables us to compare 
anisotropies of the microscopic tensor $ \vect{t} $
and the macroscopic tensor $ \bra \tilde{\vect{t}} \ket = G \vect{t} $.
We found that in the anisotropic diagram
there is a simple geometric relation between
the microscopic tensor and the corresponding macroscopic tensor.
In Sect.~\ref{sect5} we will prove some theorems to ensure the micro-macro relation.
This section is a highlight of this paper.

In Sect.~\ref{sect6} we will show simple applications of our method.
In Sect.~\ref{sect7} we restrict our consideration to molecules which have
the $ D_{2h} $ symmetry.
Then, we will reproduce the four order parameters of Straley.
In Sect.~\ref{sect8} we will give a general prescription
to formulate the Landau-de Gennes free energy model.
There we refer to the theorem which tells
a complete set of ingredients of the Landau-de Gennes free energy.
In the appendix we prove the theorem.
A real molecule may have various tensorial quantities
which are not simultaneously diagonalizable.
Our prescription is applicable even to such a general system.
Finally, we apply our prescription and 
obtain a complete Landau-de Gennes free energy for the $ D_{2h} $-symmetric molecules.
Sect.~\ref{sect9} is devoted to concluding remarks.

We would like to emphasize that our method for characterizing anisotropies 
is applicable to a general system in which
molecules may have arbitrary shapes and arbitrary tensorial quantities.
Our method is systematic and unambiguous.
The anisotropy diagram will help 
both qualitative and quantitative understandings of anisotropies.
Our prescription for formulating the Landau-de Gennes free energy
enables us to construct a complete invariant polynomial
which contains neither too many nor too few terms.

\section{Geometric order parameter}\label{sect2}
In this section we introduce the geometric order parameter.
Although it is just a modified version of de Gennes' ordering matrix,
it will give a clear and new interpretation of the ordering matrix.

Assume that a molecule has an intrinsic vectorial quantity 
$ \vect{v} = ( v_1, v_2, v_3 ) $,
which can be, for example, an electric dipole moment.
When the molecule rotates,
the vector $ \vect{v} $ is transformed to
\begin{equation}
	\tilde{\vect{v}} = R \vect{v},
	\qquad \mbox{or} \qquad
	\tilde{v}_a = \sum_{i=1}^3 R_{ai} \, v_i
\end{equation}
by a three-dimensional orthogonal matrix $ R \in SO(3) $.
The matrix elements $ R_{ai} $ satisfy
$ \sum_{a} R_{ai} R_{aj} = \delta_{ij} $ and
$ \sum_{i} R_{ai} R_{bi} = \delta_{ab} $.
Each molecule can be transformed by a different rotation matrix.
Since liquid crystal is an ensemble of molecules,
the quantity observed in the laboratory is the average
\begin{equation}
	\bra \tilde{\vect{v}} \ket = \bra R \ket \vect{v},
	\qquad \mbox{or} \qquad
	\bra \tilde{v}_a \ket = \sum_{i=1}^3 \bra R_{ai} \ket v_i.
\end{equation}
Once we know the matrix elements $ \bra R_{ai} \ket $,
we can calculate the average 
$ \bra \tilde{\vect{w}} \ket = \bra R \ket \vect{w} $
for any vectorial quantity $ \vect{w} $ of the molecule.
Most of liquid crystals have no polarity
and hence $ \bra R_{ai} \ket $ are usually zero.

Next, assume that the molecule has an intrinsic tensorial quantity 
$ \vect{t} = ( t_{ij} ) $,
which may be a dielectric susceptibility or an electric quadrupole moment.
When the molecule rotates, the tensor $ \vect{t} $ is transformed to
\begin{equation}
	\tilde{\vect{t}} = (R \otimes R) \vect{t},
	\qquad \mbox{or} \qquad
	\tilde{t}_{ab} = \sum_{i,j=1}^3 R_{ai} \, R_{bj} \, t_{ij}.
\end{equation}
Any tensor $ \vect{t} $ can be decomposed into
the scalar component, the antisymmetric component,
and the traceless symmetric component as
\begin{equation}
	t_{ij}
	= \bigg[ \frac{1}{3} \delta_{ij} \sum_{k=1}^3 t_{kk} \bigg]
	+ \bigg[ \frac{1}{2} ( t_{ij} - t_{ji} ) \bigg]
	+ \bigg[
	\frac{1}{2} ( t_{ij} + t_{ji} ) - \frac{1}{3} \delta_{ij} \sum_{k=1}^3 t_{kk}
	\bigg].
\end{equation}
If the tensor $ \vect{t} $ is traceless and symmetric,
the transformed tensor $ \tilde{\vect{t}} $ is also traceless and symmetric.
Hence we can write the components of $ \tilde{\vect{t}} $ as
\begin{equation}
	\tilde{t}_{ab}
	= 
	\frac{1}{2} ( \tilde{t}_{ab} + \tilde{t}_{ba} ) 
	- \frac{1}{3} \delta_{ab} \sum_{c=1}^3 \tilde{t}_{cc}
	=
	\sum_{i,j=1}^3
	\bigg[
	\frac{1}{2} ( R_{ai} \, R_{bj} + R_{bi} \, R_{aj} )
	- \frac{1}{3} \delta_{ab} \, \delta_{ij}
	\bigg]
	t_{ij}.
\end{equation}
Thus, the transformation law of traceless symmetric tensors is described 
as
\begin{equation}
	\tilde{t}_{ab} = \sum_{i,j=1}^3 Q_{abij} \, t_{ij}
\end{equation}
with the symmetrized traceless matrix
\begin{equation}
	Q_{abij} 
	=
	\frac{1}{2} ( R_{ai} \, R_{bj} + R_{bi} \, R_{aj} )
	- \frac{1}{3} \delta_{ab} \, \delta_{ij}.
	\label{reduced matrix}
\end{equation}
Then the average, which is an observable at a macroscopic scale, is given by
\begin{equation}
	\bra \tilde{\vect{t}} \ket = \bra Q \ket \vect{t} = G \vect{t},
	\qquad \mbox{or} \qquad
	\bra \tilde{t}_{ab} \ket 
	= \sum_{i,j=1}^3 \bra Q_{abij} \ket t_{ij}
	= \sum_{i,j=1}^3 G_{abij} \, t_{ij}.
	\label{micro-macro}
\end{equation}
The defining equation of $ \bra Q_{abij} \ket = G_{abij} $ is Eq.~(\ref{ordering matrix}).
Once we know the matrix elements $ \bra Q_{abij} \ket $,
we can calculate the average
$ \bra \tilde{\vect{u}} \ket = \bra Q \ket \vect{u} $
for any tensorial quantity $ \vect{u} $ of the molecule.
It is not necessary that
the tensors $ \vect{t} $ and $ \vect{u} $ have common principal axes.
We call $ \bra R \ket $ and $ \bra Q \ket $ 
the {\it geometric order parameters}.
More specifically,
we may call $ \bra R \ket $ 
the geometric order parameter for vectors
while we call $ \bra Q \ket $ 
the geometric order parameter for traceless symmetric tensors.
In our approach, the macroscopic observable $ \bra \tilde{\vect{t}} \ket $
is calculated as a product 
$ \bra \tilde{\vect{t}} \ket = \bra Q \ket \vect{t} $
of the geometric order parameter $ \bra Q \ket $
with the molecular intrinsic quantity $ \vect{t} $.
In this treatment we can analyze anisotropies of 
$ \vect{t} $ and $ \bra \tilde{\vect{t}} \ket $ separately.

Superficially the ordering matrix $ G_{abij} = \bra Q_{abij} \ket $
has $ 3^4 = 81 $ components but actually it has only 25 independent components~\cite{Straley1974}.
The geometric order parameter $ G $ 
transforms a traceless symmetric tensor $ \vect{t} $
into another traceless symmetric tensor $ \bra \tilde{\vect{t}} \ket = G \vect{t} $.
The set of all traceless symmetric tensors forms a 5-dimensional vector space
and $ G $ is a linear transformation of the space of traceless symmetric tensors.
Hence the ordering matrix $ G $ has $ 5^2 = 25 $ independent components.
This fact can be verified also by counting independent components of $ G_{abij} $
which are restricted by the traceless and symmetry conditions
\begin{equation}
	\sum_{a=1}^3 G_{aaij} = 0,
	\qquad
	\sum_{i=1}^3 G_{abii} = 0,
	\qquad
	G_{abij} =
	G_{baij} =
	G_{abji}.
\end{equation}

We would like to have a representation of the geometric order parameter
in which only independent components appear explicitly.
For this purpose
we will introduce the {\it reduced ordering matrix} $ \hat{G}_{\mu \nu} $ in the following.
First, we define an inner product of two tensors $ \vect{t} $ and $ \vect{u} $ as
\begin{equation}
	\bra \vect{t}, \vect{u} \ket
	=
	\tr ( \vect{t}^T \vect{u} )
	=
	\sum_{i,j=1}^3 t_{ij} \, u_{ij}.
	\label{inner product}
\end{equation}
Here $ \vect{t}^T $ is the transposition of $ \vect{t} $.
It is allowed to make a product $ \vect{t} \vect{u} $ of two tensors as
\begin{equation}
	( \vect{t} \vect{u} )_{ik}
	=
	\sum_{j=1}^3 t_{ij} \, u_{jk}.
	\label{product}
\end{equation}
Second, we introduce a basis $ \{ \vect{\xi}_1, \cdots, \vect{\xi}_5 \} $ 
of the space of traceless symmetric tensors,
\begin{eqnarray}
&&	\vect{\xi}_1 = \frac{1}{\sqrt{2}}
	\begin{pmatrix}
	0 & 1 & 0 \\
	1 & 0 & 0 \\
	0 & 0 & 0 \\
	\end{pmatrix},
	\qquad
	\vect{\xi}_2 = \frac{1}{\sqrt{2}}
	\begin{pmatrix}
	0 & 0 & 1 \\
	0 & 0 & 0 \\
	1 & 0 & 0 \\
	\end{pmatrix},
	\qquad
	\vect{\xi}_3 = \frac{1}{\sqrt{2}}
	\begin{pmatrix}
	0 & 0 & 0 \\
	0 & 0 & 1 \\
	0 & 1 & 0 \\
	\end{pmatrix},
\nonumber \\
&&
	\vect{\xi}_4 = \frac{1}{\sqrt{2}}
	\begin{pmatrix}
	1 & 0 & 0 \\
	0 &-1 & 0 \\
	0 & 0 & 0 \\
	\end{pmatrix},
	\qquad
	\vect{\xi}_5 = \frac{1}{\sqrt{6}}
	\begin{pmatrix}
	-1 & 0 & 0 \\
	0 & -1 & 0 \\
	0 & 0 & 2 \\
	\end{pmatrix}.
	\label{xis}
\end{eqnarray}
They satisfy 
$ \bra \vect{\xi}_\mu, \vect{\xi}_\nu \ket = \delta_{\mu \nu} $
with respect to the inner product (\ref{inner product}).
We write the components of $ \vect{\xi}_\mu $ as $ \xi_{\mu,ij} $ 
with indices $ \mu = 1, \cdots, 5 $ and $ i,j=1,2,3 $.
An arbitrary traceless symmetric tensor $ \vect{t} $ can be expressed
as a linear combination of
$ \{ \vect{\xi}_1, \cdots, \vect{\xi}_5 \} $,
\begin{equation}
	\vect{t} = \sum_{\mu=1}^5 c_\mu \vect{\xi}_\mu
	\label{linear combination}
\end{equation}
with the coefficients $ c_\mu = \bra \vect{\xi}_\mu, \vect{t} \ket $.
Finally, we define a 5-dimensional matrix 
$ \hat{G} = ( \hat{G}_{\mu \nu} ) $ by
\begin{equation}
	G \vect{\xi}_{\nu} 
	= \sum_{\mu = 1}^5 \vect{\xi}_{\mu} \hat{G}_{\mu \nu}, 
	\qquad
	\nu = 1, \cdots, 5.
\end{equation}
These matrix elements $ \hat{G}_{\mu \nu} $ can be calculated as
\begin{equation}
	\hat{G}_{\mu \nu} 
	=
	\bra \vect{\xi}_{\mu}, G \vect{\xi}_{\nu} \ket
	=
	\! \sum_{a,b,i,j=1,2,3} \!\!
	{\xi}_{\mu,ab} \, G_{abij} \, {\xi}_{\nu,ij}.
	\label{reduced ordering matrix}
\end{equation}
We call the 5-dimensional matrix $ \hat{G} = ( \hat{G}_{\mu \nu} ) $
the {\it reduced ordering matrix}.
By the definition it has $ 5^2 = 25 $ independent components.
When the macroscopic tensor $ \bra \tilde{\vect{t}} \ket = G \vect{t} $
is expanded in the basis as
$ \bra \tilde{\vect{t}} \ket = \sum_{\mu=1}^5 d_\mu \vect{\xi}_\mu $,
its components are given by
\begin{equation}
	d_{\mu} = \sum_{\nu =1}^5 \hat{G}_{\mu \nu} \, c_{\nu}.
\end{equation}
The components $ G_{abij} $ of the original geometric order parameter can be reconstructed 
from the components $ \hat{G}_{\mu \nu} $ of the reduced ordering matrix as
\begin{equation}
	G_{abij} 
	=
	\sum_{\mu, \nu =1}^5 
	{\xi}_{\mu,ab} \, \hat{G}_{\mu \nu} \, {\xi}_{\nu,ij}.
\end{equation}
Therefore the reduced ordering matrix 
contains the same information as the geometric order parameter.

To read out the implication of the geometric order parameter
we apply the singular value decomposition on it.
Here we review the definition of the singular value decomposition of a matrix.
For a matrix $ \hat{G} = ( \hat{G}_{\mu \nu} ) $ if a set of vectors
$ \vect{c}_\alpha = ( c_{\mu \alpha} ) $,
$ \vect{d}_\alpha = ( d_{\mu \alpha} ) $
and real numbers $ \sigma_\alpha $
$ ( \mu, \nu, \alpha = 1, \cdots, 5 ) $ satisfy
\begin{equation}
	\sum_{\nu =1}^5 \hat{G}_{\mu \nu} c_{\nu \alpha} = \sigma_\alpha d_{\mu \alpha},
	\quad
	\sum_{\mu =1}^5 \hat{G}_{\mu \nu} d_{\mu \alpha} = \sigma_\alpha c_{\nu \alpha},
	\quad
	\sum_{\mu =1}^5 c_{\mu \alpha} c_{\mu \beta} = \delta_{\alpha \beta},
	\quad
	\sum_{\mu =1}^5 d_{\mu \alpha} d_{\mu \beta} = \delta_{\alpha \beta},
	\label{singular equation}
\end{equation}
then the vector $ \vect{c}_\alpha $ is called the right singular vector,
$ \vect{d}_\alpha $ is called the left singular vector,
and $ \sigma_\alpha $ is called the singular value.
The above equations can be written more concisely as
\begin{equation}
	\hat{G} \vect{c}_\alpha = \sigma_\alpha \vect{d}_\alpha,
	\qquad
	\hat{G}^T \vect{d}_\alpha = \sigma_\alpha \vect{c}_\alpha,
	\qquad
	\bra \vect{c}_\alpha, \vect{c}_\beta \ket = \delta_{\alpha \beta},
	\qquad
	\bra \vect{d}_\alpha, \vect{d}_\beta \ket = \delta_{\alpha \beta}.
	\label{SVD}
\end{equation}
Here $ \hat{G}^T $ is the transposed matrix of $ \hat{G} $.
It is always possible to make $ \sigma_\alpha $ non-negative
by choosing $ \vect{c}_\alpha $ and $ \vect{d}_\alpha $ suitably.
For a symmetric matrix $ \hat{G} = \hat{G}^T $,
the left singular vector and the right singular vector coincide and
they are called an eigenvector. 
In this case the singular value is called an eigenvalue.
{}From the singular vectors and singular values we can construct matrices
\begin{equation}
	C = ( c_{\mu \alpha} ),
	\qquad
	D = ( d_{\mu \alpha} ),
	\qquad
	\Sigma = ( \sigma_\alpha  \delta_{\alpha \beta} ).
\end{equation}
Note that $ \Sigma $ is a diagonal matrix.
Then the set of equations (\ref{singular equation}) is equivalent to
\begin{equation}
	\hat{G} C = D \Sigma,
	\qquad
	D^T \hat{G} = \Sigma C^T,
	\qquad
	C^T C = I,
	\qquad
	D^T D = I,
\end{equation}
which implies $ D^T \hat{G} C = \Sigma $. 
This is a generalization of diagonalization of a matrix.
It can be rewritten as
\begin{equation}
	\hat{G} = D \Sigma C^T
\end{equation}
and this expression is called the singular value decomposition of $ \hat{G} $.

Now we apply the singular value decomposition to the reduced ordering matrix $ \hat{G} $
to understand the implication of the geometric order parameter.
Once we know the singular vectors of $ \hat{G} $,
$ \vect{c}_\alpha = ( c_{\mu \alpha} ) $ and
$ \vect{d}_\alpha = ( d_{\mu \alpha} ) $,
we can construct tensors
\begin{equation}
	\vect{t}_\alpha =
	\sum_{\mu=1}^5  \vect{\xi}_\mu c_{\mu \alpha},
	\qquad
	\vect{u}_\alpha =
	\sum_{\mu=1}^5  \vect{\xi}_\mu d_{\mu \alpha}.
\end{equation}
Then the definition of singular vectors (\ref{SVD}) implies
\begin{equation}
	G \vect{t}_\alpha = \sigma_\alpha \vect{u}_\alpha,
	\qquad
	\bra \vect{t}_\alpha, \vect{t}_\beta \ket = \delta_{\alpha \beta},
	\qquad
	\bra \vect{u}_\alpha, \vect{u}_\beta \ket = \delta_{\alpha \beta}.
	\label{singular}
\end{equation}
On the other hand,
as discussed at (\ref{micro-macro}),
when the molecule has an intrinsic physical quantity $ \vect{t}_\alpha $,
the average $ \bra \tilde{\vect{t}}_\alpha \ket = G \vect{t}_\alpha $ 
will be observed by a macroscopic measurement.
The observed value is now given as
$ \bra \tilde{\vect{t}}_\alpha \ket = \sigma_\alpha \vect{u}_\alpha $.
The coefficient $ \sigma_\alpha $ takes its value in the range
$ 0 \le \sigma_\alpha \le 1 $
and is called the {\it degree of order} or the {\it strength of realization}.
The reason why $ \sigma_\alpha $ is in the range $ 0 \le \sigma_\alpha \le 1 $
will be explained in Sect.~\ref{sect5} as a corollary of theorem 1.
The tensor $ \vect{t}_\alpha $ is called the {\it microscopic singular tensor}
and 
$ \vect{u}_\alpha $ is called the {\it macroscopic singular tensor}.
It is convenient to arrange them in the order
$ \sigma_1 \ge \sigma_2 \ge \cdots \ge \sigma_5 $.
Then, if each molecule has a quantity represented by $ \vect{t}_\alpha $,
the ensemble of molecules exhibits the quantity $ \vect{u}_\alpha $ at the macroscopic scale
with the strength $ \sigma_\alpha $.
If $ \sigma_\alpha = 0 $, the effect of the molecular quantity $ \vect{t}_\alpha $ 
disappears at the macroscopic scale.

Let us summarize the above discussion.
The equation (\ref{micro-macro}) relates 
the microscopic tensorial quantity $ \vect{t} $
to the macroscopic observable $ \bra \tilde{\vect{t}} \ket $.
The equation $ \bra \tilde{\vect{t}} \ket = G \vect{t} $
can be rewritten symbolically as
\begin{equation}
	\mbox{(macroscopic observable)}
	=
	\mbox{(geometric order parameter)} 
	\times
	\mbox{(microscopic quantity)}.
\end{equation}
Furthermore, the equation (\ref{singular})
tells that the molecular quantity $ \vect{t}_\alpha $ manifests
itself as the macroscopic quantity $ \vect{u}_\alpha $
with the strength $  \sigma_\alpha $.
This relation 
$ G \vect{t}_\alpha = \sigma_\alpha \vect{u}_\alpha $
can be expressed symbolically as
\begin{eqnarray}
&&	\mbox{(geometric order parameter)} 
	\times
	\mbox{(microscopic singular tensor)}
	\nonumber \\
&&	\quad =
	\mbox{(strength of realization)} 
	\times
	\mbox{(macroscopic singular tensor)}.
\end{eqnarray}
In this way we can read the implication of the geometric order parameter $ G $.

We would like to mention another interesting property of the geometric parameters.
In a nematic phase orientations of molecules are fluctuating.
The orientation of each molecule is specified with a three-dimensional rotation matrix
$ R \in SO(3) $.
Then distribution of the molecular orientations is described 
by a probability density function $ p (R) $ over $ SO(3) $
and the average of a physical quantity $ f(R) $ which depends on the orientation of a molecule
is given by the integral
\begin{equation}
	\bra f \ket 
	= \int f(R) \, p(R) \, dR
	= \frac{1}{8 \pi^2}
	\int f(R) \, p(R) \, \sin \theta \, d \theta \, d \phi \, d \psi.
\end{equation}
In the last line we used the Euler angles $ ( \theta, \phi, \psi ) $
to specify the rotation matrix $ R $.
Note that $ Q_{abij} (R) $ defined in (\ref{reduced matrix}) is 
a function of $ R \in SO(3) $. Furthermore, if we define
\begin{equation}
	\hat{Q}_{\mu \nu} 
	=
	\bra \vect{\xi}_{\mu}, Q \vect{\xi}_{\nu} \ket
	=
	\! \sum_{a,b,i,j=1,2,3} \!\!
	{\xi}_{\mu,ab} \, Q_{abij} \, {\xi}_{\nu,ij},
\end{equation}
$ \hat{Q}_{\mu \nu}(R) $ is also a function of $ R \in SO(3) $.
The matrix $ \hat{Q} (R) = ( \hat{Q}_{\mu \nu} (R) ) $
forms a 5-dimensional irreducible representation of the rotation group $ SO(3) $.
Namely, it satisfies $ \hat{Q} (R R') = \hat{Q} (R) \hat{Q} (R') $
for any $ R, R' \in SO(3) $.
If we know the probability density $ p(R) $,
we can calculate the averages 
$ \bra R_{ai} \ket $ and $ \bra \hat{Q}_{\mu \nu} \ket $.
Actually, the inverse of this statement holds.
Once we know the averages
$ \bra R_{ai} \ket $ and $ \bra \hat{Q}_{\mu \nu} \ket $,
we can determine the probability density $ p(R) $ via
\begin{equation}
	p(R) = 1 
	+ 3 \sum_{a,i=1}^3 
	\bra R_{ai} \ket \, R_{ai}
	+ 5 \sum_{\mu, \nu=1}^5 
	\bra \hat{Q}_{\mu \nu} \ket \, \hat{Q}_{\mu \nu}
	+ \cdots.
\end{equation}
This equation is regarded as an expansion of $ p(R) $ 
in powers of $ R_{ai} $.
It is easily proved by applying the Peter-Weyl theorem~\cite{Kobayashi1999}
of group representation theory.
In this way
the geometric order parameters 
completely characterize the geometric and statistical properties
of the nematic phase.

\section{Elementary attempts to characterize anisotropy}\label{sect3}
In the previous section we argued that
the microscopic tensorial quantity $ \vect{t} $ is related to 
the macroscopic tensorial quantity $ \bra \tilde{\vect{t}} \ket $
via the geometric order parameter $ G $ 
as $ \bra \tilde{\vect{t}} \ket = G  \vect{t} $.
We also showed that the implication of the geometric order parameter $ G $
can be analyzed via the singular value decomposition.
The remaining problem is to provide a systematic method to analyze
properties of tensorial quantities, $ \vect{t} $ or $ \bra \tilde{\vect{t}} \ket $,
particularly their anisotropy.
This is the subject we will discuss in this section.

Here we discuss briefly some attempts to characterize anisotropy
of a symmetric tensor $ \vect{t} = ( t_{ij} ) $ (the trace is not necessarily zero).
The tensor has three principal axes and three eigenvalues $ \lambda_1, \lambda_2, \lambda_3 $.
By choosing the spacial coordinate suitably,
we can transform it in a diagonal form
\begin{equation}
	\vect{t} =
	\begin{pmatrix}
	\lambda_1 & 0 & 0 \\
	0 & \lambda_2 & 0 \\
	0 & 0 & \lambda_3 
	\end{pmatrix}.
\end{equation}
When the three eigenvalues coincide, it is said that the tensor is {\it isotropic}.
When two of the three eigenvalues coincide, the tensor is {\it uniaxial}.
When the three are distinct, the tensor is {\it biaxial}.
We would like to define indices which indicate quantitatively
the degree of anisotropy and the degree of biaxiality.

As a candidate for the index of anisotropy we may introduce
\begin{equation}
	\Delta = 
	( \lambda_1 - \lambda_2 )^2 +
	( \lambda_2 - \lambda_3 )^2 +
	( \lambda_3 - \lambda_1 )^2.
	\label{Delta}
\end{equation}
It is obvious that $ \Delta $ is non-negative.
If and only if $ \Delta = 0 $, the tensor is isotropic.
On the other hand, we define the average of the eigenvalues
$ m = \frac{1}{3} ( \lambda_1 + \lambda_2 + \lambda_3 ) $
and the standard deviation
\begin{equation}
	\tilde{\Delta} = \frac{1}{3}
	\Big\{
	( \lambda_1 - m )^2 +
	( \lambda_2 - m )^2 +
	( \lambda_3 - m )^2 
	\Big\}.
	\label{tildeDelta}
\end{equation}
We may take $ \tilde{\Delta} $ as another index of anisotropy
but actually they are related as
\begin{equation}
	\Delta = 9 \, \tilde{\Delta}.
\end{equation}
Hence, $ \tilde{\Delta} $ differs from $ \Delta $ only by a coefficient.

As a candidate for the index of biaxiality we may introduce
\begin{equation}
	\kappa = 
	\Big\{
	( \lambda_1 - \lambda_2 )
	( \lambda_2 - \lambda_3 )
	( \lambda_3 - \lambda_1 ) \Big\}^2.
	\label{kappa}
\end{equation}
The index $ \kappa $ is non-negative.
It is obvious that
the tensor is biaxial if and only if $ \kappa \ne 0 $.
The index $ \kappa $ is called the discriminant 
in the context of theory of algebraic equations.
We explain this point briefly.
The eigenvalues of the matrix $ \vect{t} = ( t_{ij} ) $ 
are roots of the cubic equation
\begin{equation}
	\det ( x I - \vect{t} )
	=
	x^3 + a x^2 + b x + c
	=
	0.
	\label{cubic equation}
\end{equation}
The coefficients and roots are related as
\begin{eqnarray}
&&	a 
	= - ( \lambda_1 + \lambda_2 + \lambda_3 )
	= - \tr \, \vect{t},
	\\
&&	b 
	= \lambda_1 \lambda_2 + \lambda_2 \lambda_3 + \lambda_3 \lambda_1 
	= \frac{1}{2} ( \lambda_1 + \lambda_2 + \lambda_3 )^2
	- \frac{1}{2} ( \lambda_1^2 + \lambda_2^2 + \lambda_3^2 )
	= \frac{1}{2} (\tr \, \vect{t})^2 - \frac{1}{2} \tr (\vect{t}^2),
	\\
&&	c
	= - \lambda_1 \lambda_2 \lambda_3
	= - \det \vect{t}.
\end{eqnarray}
$ \kappa = 0 $ is a necessary and sufficient condition for existence of a multiple root.
It is known that the discriminant is expressed in terms of the coefficients as
\begin{equation}
	\kappa = a^2 b^2 + 18 abc - 4 b^3 - 4 a^3 c -27 c^2.
\end{equation}
Similarly, the degree of anisotropy (\ref{Delta})
is expressed in terms of the matrix $ \vect{t} $ as
\begin{eqnarray}
	\Delta 
&=& 
	( \lambda_1 - \lambda_2 )^2 +
	( \lambda_2 - \lambda_3 )^2 +
	( \lambda_3 - \lambda_1 )^2
	\nonumber \\
&=&	2 ( \lambda_1^2 + \lambda_2^2 + \lambda_3^2 )
	-2 ( \lambda_1 \lambda_2 + \lambda_2 \lambda_3 + \lambda_3 \lambda_1 )
	\nonumber \\
&=&	2 \, \tr ( \vect{t}^2 )
	- \{ (\tr \, \vect{t})^2 - \tr (\vect{t}^2) \}
	\nonumber \\
&=&	3 \, \tr ( \vect{t}^2 ) - (\tr \, \vect{t})^2.
\end{eqnarray}

We may use $ \Delta $ and $ \kappa $ as indices of anisotropy and biaxiality.
But, particularly, $ \kappa $ is not convenient for calculation.
What is worse, the index $ \kappa $ is not useful for comparing 
the biaxiality of the microscopic tensor $ \vect{t} $
with the biaxiality of the macroscopic tensor $ \bra \tilde{\vect{t}} \ket $.
In the next section we will introduce a more convenient method
to evaluate and classify anisotropies.

\section{Anisotropy diagram}\label{sect4}
Here we will introduce a diagrammatic method to characterize anisotropy of a given tensor.
Our diagram will be convenient for comparing anisotropies of various tensors.
It will be shown that the microscopic tensor and the macroscopic tensor
have a definite relation in our diagram.

In the following any tensor is assumed to be traceless and symmetric.
The eigenvalues of a tensor $ \vect{t} $ are denoted as
$ \lambda_1, \lambda_2, \lambda_3 $.
It is a usual convention to arrange the eigenvalues in the order
$ \lambda_3 \ge \lambda_1 \ge \lambda_2 $.
Under the assumption of tracelessness
the sum of the three eigenvalues is zero.
Here we give definitions for classification of traceless symmetric tensors.
The tensor is {\it isotropic} if $ \vect{t} = 0 $.
Otherwise, it is {\it anisotropic}.
When two of the three eigenvalues coincide, the tensor is {\it uniaxial}.
Moreover, when a uniaxial tensor satisfies $ \det \vect{t} > 0 $,
namely, $ \lambda_3 > 0 > \lambda_1 = \lambda_2 $,
it is said that the tensor has {\it positive uniaxiality}.
A positively uniaxial tensor has eigenvalues
$ ( \lambda_1, \lambda_2, \lambda_3 ) = \lambda ( -1, -1, 2) $
with a positive coefficient $ \lambda $.
On the other hand, when a uniaxial tensor satisfies $ \det \vect{t} < 0 $,
namely, $ \lambda_3 = \lambda_1 > 0 > \lambda_2 $,
it is said that the tensor has {\it negative uniaxiality}.
A negatively uniaxial tensor has eigenvalues
$ ( \lambda_1, \lambda_2, \lambda_3 ) = \lambda ( 1, -2, 1) $
with a positive coefficient $ \lambda $.
When $ \det \vect{t} = 0 $ and $ \vect{t} \ne 0 $,
namely, $ \lambda_2 = - \lambda_3 $ and $ \lambda_1 = 0 $,
it is said that the tensor has {\it maximal biaxiality}.
A maximally biaxial tensor has eigenvalues
$ ( \lambda_1, \lambda_2, \lambda_3 ) = \lambda ( 0, -1, 1) $
with a positive coefficient $ \lambda $.
\begin{figure}[bt]
\begin{center}
\scalebox{0.45}{
\includegraphics{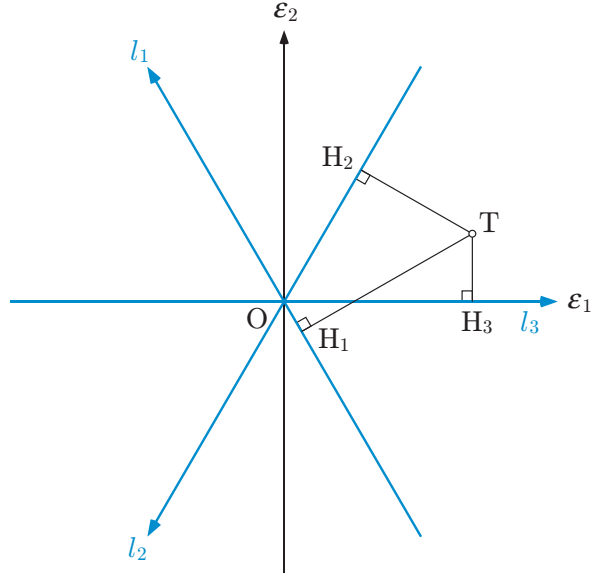}
}
\end{center}
\vspace{-2mm}
\caption{\label{fig1}
In the anisotropy diagram a tensor 
$ \vect{t} = \mbox{diag} ( \lambda_1, \lambda_2, \lambda_3 ) $
is represented by a point T $ = ( \varepsilon_1, \varepsilon_2 ) $.
The lengths of segments $ ( \mbox{OH}_1, \mbox{OH}_2, \mbox{OH}_3 ) $
are equal to $ \sqrt{\frac{3}{2}} ( \lambda_1, \lambda_2, \lambda_3 ) $.
}
\end{figure}

Any traceless symmetric tensor can be diagonalized and parameterized in the form
\begin{equation}
	\vect{t} =
	\begin{pmatrix}
	\lambda_1 & 0 & 0 \\
	0 & \lambda_2 & 0 \\
	0 & 0 & \lambda_3
	\end{pmatrix}
	= \varepsilon_1 \frac{1}{\sqrt{6}}
	\begin{pmatrix}
	-1& 0 & 0 \\
	0 &-1 & 0 \\
	0 & 0 & 2
	\end{pmatrix}
	+ \varepsilon_2 \frac{1}{\sqrt{2}}
	\begin{pmatrix}
	1 & 0 & 0 \\
	0 &-1 & 0 \\
	0 & 0 & 0
	\end{pmatrix}.
	\label{tensor coordinate}
\end{equation}
The coefficients $ ( \varepsilon_1, \varepsilon_2 ) $
are the same things as $ ( c_5, c_4 ) $ in (\ref{linear combination})
and they are related to the eigenvalues as
\begin{eqnarray}
&&	\lambda_1 
	= - \frac{1}{\sqrt{6}} \, \varepsilon_1 + \frac{1}{\sqrt{2}} \, \varepsilon_2
	= \sqrt{\frac{2}{3}} \cdot \frac{1}{2}
	( -1,   \sqrt{3} ) 
	\begin{pmatrix} \varepsilon_1 \\ \varepsilon_2 \end{pmatrix},
	\label{co1}
	\\
&&	\lambda_2 
	= - \frac{1}{\sqrt{6}} \, \varepsilon_1 - \frac{1}{\sqrt{2}} \, \varepsilon_2
	= \sqrt{\frac{2}{3}} \cdot \frac{1}{2}
	( -1, - \sqrt{3} ) 
	\begin{pmatrix} \varepsilon_1 \\ \varepsilon_2 \end{pmatrix},
	\label{co2}
	\\
&&	\lambda_3 
	= \sqrt{\frac{2}{3}} \, \varepsilon_1
	= \sqrt{\frac{2}{3}} \, (1,0) 
	\begin{pmatrix} \varepsilon_1 \\ \varepsilon_2 \end{pmatrix},
	\label{co3}
\end{eqnarray}
or inversely 
\begin{eqnarray}
&&	\varepsilon_1 =\sqrt{\frac{3}{2}} \, \lambda_3,
	\label{varepsilon_1}
	\\
&&	\varepsilon_2 
	= \frac{1}{\sqrt{2}} ( \lambda_1 - \lambda_2 )
	= \frac{1}{\sqrt{2}} ( 2 \lambda_1 + \lambda_3 )
	= \frac{1}{\sqrt{2}} (-2 \lambda_2 - \lambda_3 ),
	\label{varepsilon_2}
	\\
&&	0 = \lambda_1 + \lambda_2 + \lambda_3.
\end{eqnarray}
The inner product (\ref{inner product}) is used to define the norm of the tensor
\begin{equation}
	\alpha
	= || \vect{t} ||
	= \sqrt{ \bra \vect{t}, \vect{t} \ket }
	= \sqrt{ (\lambda_1)^2 + (\lambda_2)^2 + (\lambda_3)^2 }
	= \sqrt{ (\varepsilon_1)^2 + (\varepsilon_2)^2 }.
	\label{alpha}
\end{equation}
Then $ \alpha^2 $ is equal to the anisotropy index $ 3 \tilde\Delta $ 
which was defined at (\ref{tildeDelta}).
For the tensor $ \vect{t} $ we plot a point T 
whose Cartesian coordinate is $ ( \varepsilon_1, \varepsilon_2 ) $
as shown in Fig.~1.
Thus each point in the plane defines a corresponding traceless symmetric tensor.
This plane diagram is called an {\it anisotropy diagram}.
The value of $ \alpha $ is equal to 
the distance between the point T and the origin O of the coordinate.

We explain how to draw the anisotropy diagram in detail.
For a given traceless tensor $ \vect{t} $ one calculates the eigenvalues 
$ ( \lambda_1, \lambda_2, \lambda_3 ) $.
Next one calculates $ ( \varepsilon_1, \varepsilon_2 ) $ 
using Eqs.~(\ref{varepsilon_1}), (\ref{varepsilon_2}).
Plot a point T $ = ( \varepsilon_1, \varepsilon_2 ) $ in the Cartesian coordinate.
This is the point representing the tensor.
Draw three lines 
$ \ell_1, \ell_2, \ell_3 $
which run through the origin $ \rm{O} = (0,0) $ 
in the direction 
$ ( -\frac{1}{2},   \frac{\sqrt{3}}{2} ) $,
$ ( -\frac{1}{2}, - \frac{\sqrt{3}}{2} ) $,
$ (1,0) $,
respectively.
Draw a line $ m_1 $ which runs through the point T 
and is perpendicular to the line $ \ell_1 $.
The intersection of $ \ell_1 $ and $ m_1 $ is denoted as $ \rm{H}_1 $.
Similarly, draw lines $ m_2 $ and $ m_3 $ which run through the point T 
and are perpendicular to the line $ \ell_2 $ and $ \ell_3 $, respectively.
The intersection of $ \ell_2 $ and $ m_2 $ is denoted as $ \rm{H}_2 $.
The intersection of $ \ell_3 $ and $ m_3 $ is denoted as $ \rm{H}_3 $.
Then the lengths of $ \rm{OH}_1, \rm{OH}_2, \rm{OH}_3 $
are equal to 
$ \sqrt{\frac{3}{2}} \, \lambda_1,
  \sqrt{\frac{3}{2}} \, \lambda_2,
  \sqrt{\frac{3}{2}} \, \lambda_3 $, respectively.
In this way, 
we can determine the set of eigenvalues from the representing point,
and vice versa.

\begin{figure}[bt]
\begin{center}
\scalebox{0.45}{
\includegraphics{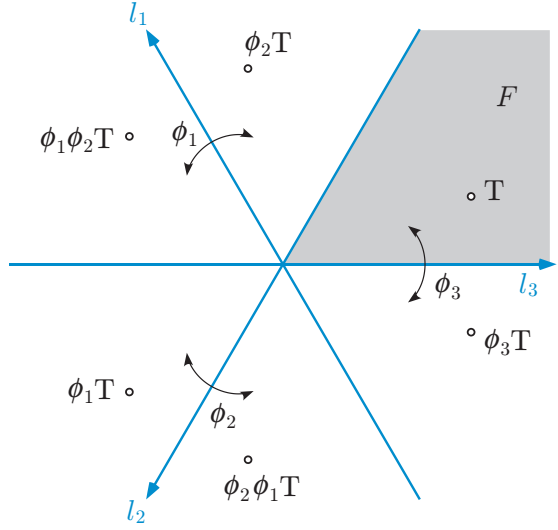}
}
\end{center}
\vspace{-2mm}
\caption{\label{fig2}
A permutation of the eigenvalues $ \lambda_2 \leftrightarrow \lambda_3 $
induces the transformation $ \phi_1 $ of the anisotropy diagram,
which is a reflection about the line $ \ell_1 $.
Similarly, other permutations 
$ \lambda_3 \leftrightarrow \lambda_1 $ and
$ \lambda_1 \leftrightarrow \lambda_2 $
induce $ \phi_2 $ and $ \phi_3 $, respectively.
The permutations of the eigenvalues generate six equivalent points.
There is a unique representative point in the fundamental domain $ F $.
}
\end{figure}
If it is not requested to arrange 
the eigenvalues in the order $ \lambda_3 \ge \lambda_1 \ge \lambda_2 $
and it is allowed to rearrange them,
a point in the anisotropy diagram corresponding to the given tensor
is not unique.
The operation exchanging $ \lambda_2 \leftrightarrow \lambda_3 $
induces a transformation of the coordinate of the anisotropy diagram as
\begin{eqnarray*}
&&	\varepsilon'_1
	= \sqrt{\frac{3}{2}} \, \lambda_2
	= \sqrt{\frac{3}{2}} 
	\left( - \frac{1}{\sqrt{6}} \, \varepsilon_1 
	- \frac{1}{\sqrt{2}} \, \varepsilon_2 \right)
	= 
	- \frac{1}{2} \, \varepsilon_1 
	- \frac{\sqrt{3}}{2} \, \varepsilon_2,
	\\
&&	\varepsilon'_2 
	= \frac{1}{\sqrt{2}} (  2 \lambda_1 + \lambda_2 )
	=  \frac{1}{\sqrt{2}} 
	\left( 
	\frac{-3}{\sqrt{6}} \, \varepsilon_1 + \frac{1}{\sqrt{2}} \, \varepsilon_2
	\right)
	=  
	- \frac{\sqrt{3}}{2} \, \varepsilon_1 + \frac{1}{2} \, \varepsilon_2,
\end{eqnarray*}
which can be summarized as
\begin{equation}
	\phi_1 : 
	\begin{pmatrix}
	\varepsilon_1 \\ \varepsilon_2 
	\end{pmatrix}
	\mapsto
	\begin{pmatrix}
	\varepsilon_1' \\
	\varepsilon_2' 
	\end{pmatrix}
	= \frac{1}{2}
	\begin{pmatrix}
	-1 & - \sqrt{3} \\
	- \sqrt{3} & 1
	\end{pmatrix}
	\begin{pmatrix}
	\varepsilon_1 \\
	\varepsilon_2 
	\end{pmatrix}.
	\label{phi_1}
\end{equation}
The permutation $ \lambda_3 \leftrightarrow \lambda_1 $
induces a transformation
\begin{eqnarray*}
&&	\varepsilon'_1 
	= \sqrt{\frac{3}{2}} \, \lambda_1
	= \sqrt{\frac{3}{2}} 
	\left(
	- \frac{1}{\sqrt{6}} \, \varepsilon_1 + \frac{1}{\sqrt{2}} \, \varepsilon_2
	\right)
	= 
	- \frac{1}{2} \, \varepsilon_1 + \frac{\sqrt{3}}{2} \, \varepsilon_2,
	\\
&&	\varepsilon_2'
	= \frac{1}{\sqrt{2}} ( - 2 \lambda_2 - \lambda_1 )
	=  \frac{1}{\sqrt{2}} 
	\left( 
	\frac{3}{\sqrt{6}} \, \varepsilon_1 + \frac{1}{\sqrt{2}} \, \varepsilon_2
	\right)
	=  
	\frac{\sqrt{3}}{2} \, \varepsilon_1 + \frac{1}{2} \, \varepsilon_2,
\end{eqnarray*}
namely,
\begin{equation}
	\phi_2 : 
	\begin{pmatrix}
	\varepsilon_1 \\ \varepsilon_2 
	\end{pmatrix}
	\mapsto
	\begin{pmatrix}
	\varepsilon_1' \\
	\varepsilon_2' 
	\end{pmatrix}
	= \frac{1}{2}
	\begin{pmatrix}
	-1 & \sqrt{3} \\
	\sqrt{3} & 1
	\end{pmatrix}
	\begin{pmatrix}
	\varepsilon_1 \\
	\varepsilon_2 
	\end{pmatrix}.
	\label{phi_2}
\end{equation}
Another permutation $ \lambda_1 \leftrightarrow \lambda_2 $ 
induces a transformation
\begin{equation}
	\phi_3 : 
	\begin{pmatrix}
	\varepsilon_1 \\ \varepsilon_2 
	\end{pmatrix}
	\mapsto
	\begin{pmatrix}
	\varepsilon'_1 \\
	\varepsilon'_2 
	\end{pmatrix}
	=
	\begin{pmatrix}
	1 & 0 \\
	0 & -1
	\end{pmatrix}
	\begin{pmatrix}
	\varepsilon_1 \\
	\varepsilon_2 
	\end{pmatrix}.
	\label{phi_3}
\end{equation}
A point T is moved to the point $ \phi_1 {\rm T} $ by the mapping $ \phi_1 $.
Furthermore, it can be moved to the point $ \phi_2 \phi_1 {\rm T} $ by $ \phi_2 $,
and so on.
In the anisotropy diagram Fig.~2,
the transformations $ \phi_1, \phi_2, \phi_3 $ are reflections
with respect to the lines $ \ell_1, \ell_2, \ell_3 $, respectively.
The set of transformations
$ \{ \phi_1, \phi_2, \phi_3 \} $
generates the third permutation group $ {\frak S}_3 $,
which has $ 3! = 6 $ elements.
Under the actions of $ {\frak S}_3 $
a generic point T in the anisotropy diagram leaves six points on its trajectory.
These trajectory points $ \rm{T}_1, \cdots, \rm{T}_6 $ are equivalent
as a representative of the tensor $ \vect{t} $.
If we impose the condition $ \lambda_3 \ge \lambda_1 $,
Eqs.~(\ref{co1}) and (\ref{co3}) imply
$ \sqrt{3} \, \varepsilon_1 \ge \varepsilon_2 $.
Moreover, if we impose the condition $ \lambda_1 \ge \lambda_2 $,
Eq.~(\ref{varepsilon_2}) implies
$ \varepsilon_2 \ge 0 $.
Hence, if the eigenvalues are arranged to satisfy the conventional ordering
$ \lambda_3 \ge \lambda_1 \ge \lambda_2 $,
a unique representative point is chosen in the domain
\begin{equation}
	F = \{ 
	( \varepsilon_1, \varepsilon_2 ) \, | \,
	0 \le \varepsilon_2 \le \sqrt{3} \, \varepsilon_1 \},
\end{equation}
which we call the {\it fundamental domain} of the anisotropy diagram.

We can use the radius $ \alpha $ and an angle $ \gamma $
to parameterize the coordinate of the anisotropy diagram as
\begin{equation}
	\varepsilon_1 = \alpha \cos \gamma, \qquad
	\varepsilon_2 = \alpha \sin \gamma.
	\label{radius-angle}
\end{equation}
By substituting these variables into (\ref{co1})-(\ref{co3}) and (\ref{kappa}), 
we obtain an expression for the index of biaxiality 
\begin{equation}
	\kappa 
	= 
	\frac{1}{2} 
	\Big[
	\varepsilon_2 \,
	\big\{ 3 (\varepsilon_1)^2 - (\varepsilon_2)^2 \big\}
	\Big]^2
	= 
	\frac{1}{2} \, \alpha^6
	\sin^2 \gamma \, ( 3 - 4 \sin^2 \gamma )^2.
	\label{kappa gamma}
\end{equation}
\begin{figure}[bt]
\begin{center}
\scalebox{0.45}{
\includegraphics{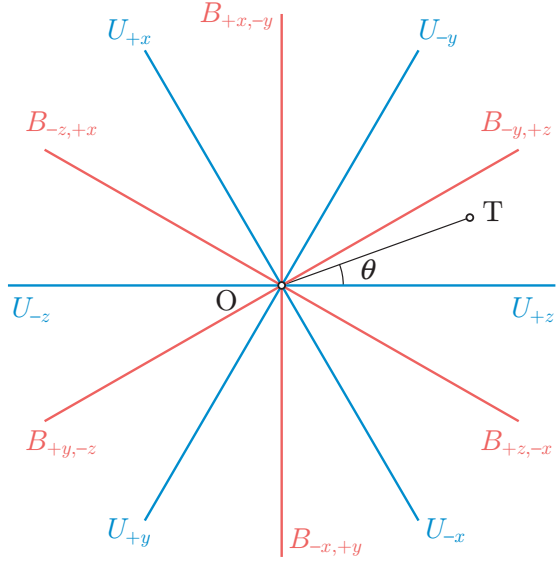}
}
\end{center}
\vspace{-2mm}
\caption{\label{fig3}
The half lines $ \{ U_j \} $ are uniaxial lines while
the half lines $ \{ B_k \} $ are maximally biaxial lines.
The degree of the anisotropy of the tensor $ \vect{t} $ is $ \alpha = $ 
the length of the segment OT.
The degree of the biaxiality is $ \beta = \frac{6}{\pi} \theta $.
}
\end{figure}
If the angle $ \gamma $ is varied,
$ \kappa $ takes the maximum value $ \frac{1}{2} \alpha^6 $
when $ \sin^2 \gamma = \frac{1}{4} $ or 1.
Hence the biaxiality becomes the maximum at
$ \gamma = 
\frac{\pi}{6}, \frac{\pi}{2}, \frac{5\pi}{6}, 
\frac{7\pi}{6}, \frac{3\pi}{2}, \frac{11\pi}{6} $.
On the other hand,
$ \kappa $ takes the minimum value $ 0 $
when $ \sin^2 \gamma = 0 $ or $ \frac{3}{4} $.
Hence the biaxiality vanishes at
$ \gamma = 
0, \frac{\pi}{3}, \frac{2 \pi}{3}, 
\pi, \frac{4 \pi}{3}, \frac{5 \pi}{3} $.
In the anisotropy diagram Fig.~3, we introduce a family of half lines
\begin{equation}
	\left\{
	\begin{array}{l}
	U_{+x} = \{ 
	( \varepsilon_1, \varepsilon_2 ) = \frac{1}{2} r (-1, \sqrt{3}) \, | \,
	r > 0 \},
	\\ \\
	U_{-x} = \{ 
	( \varepsilon_1, \varepsilon_2 ) = - \frac{1}{2} r (-1, \sqrt{3}) \, | \,
	r > 0 \},
	\\ \\
	U_{+y} = \{ 
	( \varepsilon_1, \varepsilon_2 ) = \frac{1}{2} r (-1, - \sqrt{3}) \, | \,
	r > 0 \},
	\\ \\
	U_{-y} = \{ 
	( \varepsilon_1, \varepsilon_2 ) = - \frac{1}{2} r (-1, - \sqrt{3}) \, | \,
	r > 0 \},
	\\ \\
	U_{+z} = \{ 
	( \varepsilon_1, \varepsilon_2 ) = r (1,0) \, | \,
	r > 0 \},
	\\ \\
	U_{-z} = \{ 
	( \varepsilon_1, \varepsilon_2 ) = -r (1,0) \, | \,
	r > 0 \}.
	\end{array}
	\right.
\end{equation}
These lines divide the anisotropy diagram into six domains.
Tensors which belong to 
$ U_{\pm x}, U_{\pm y}, U_{\pm z} $ are 
\begin{equation}
	\pm \frac{r}{\sqrt{6}}
	\begin{pmatrix}
	2 & 0 & 0 \\
	0 &-1 & 0 \\
	0 & 0 &-1
	\end{pmatrix},
	\qquad
	\pm \frac{r}{\sqrt{6}}
	\begin{pmatrix}
	-1& 0 & 0 \\
	0 & 2 & 0 \\
	0 & 0 &-1
	\end{pmatrix},
	\qquad
	\pm \frac{r}{\sqrt{6}}
	\begin{pmatrix}
	-1& 0 & 0 \\
	0 &-1 & 0 \\
	0 & 0 & 2
	\end{pmatrix},
\end{equation}
respectively.
The half lines $ \{ U_j \} $ are called the {\it uniaxial lines}.
Similarly, we introduce another family of half lines
\begin{equation}
	\left\{
	\begin{array}{l}
	B_{+x,-y} = \{ 
	( \varepsilon_1, \varepsilon_2 ) = r (0,1) \, | \,
	r > 0 \},
	\\ \\
	B_{-x,+y} = \{ 
	( \varepsilon_1, \varepsilon_2 ) = - r (0,1) \, | \,
	r > 0 \},
	\\ \\
	B_{+y,-z} = \{ 
	( \varepsilon_1, \varepsilon_2 ) = \frac{1}{2} r (-\sqrt{3},-1) \, | \,
	r > 0 \},
	\\ \\
	B_{-y,+z} = \{ 
	( \varepsilon_1, \varepsilon_2 ) = - \frac{1}{2} r (-\sqrt{3},-1) \, | \,
	r > 0 \},
	\\ \\
	B_{+z,-x} = \{ 
	( \varepsilon_1, \varepsilon_2 ) = \frac{1}{2} r (\sqrt{3},-1) \, | \,
	r > 0 \},
	\\ \\
	B_{-z,+x} = \{ 
	( \varepsilon_1, \varepsilon_2 ) = - \frac{1}{2} r (\sqrt{3},-1) \, | \,
	r > 0 \}.
	\end{array}
	\right.
\end{equation}
Then tensors which belong to $ B_{\pm x, \mp y}, B_{\pm y, \mp z}, B_{\pm z, \mp x} $
are
\begin{equation}
	\pm \frac{r}{\sqrt{2}}
	\begin{pmatrix}
	1 & 0 & 0 \\
	0 &-1 & 0 \\
	0 & 0 & 0
	\end{pmatrix},
	\qquad
	\pm \frac{r}{\sqrt{2}}
	\begin{pmatrix}
	0 & 0 & 0 \\
	0 & 1 & 0 \\
	0 & 0 &-1
	\end{pmatrix},
	\qquad
	\pm \frac{r}{\sqrt{2}}
	\begin{pmatrix}
	-1& 0 & 0 \\
	0 & 0 & 0 \\
	0 & 0 & 1
	\end{pmatrix}.
\end{equation}
The half lines $ \{ B_{k} \} $ are called the {\it maximally biaxial lines}.

Using the anisotropy diagram 
we can classify tensors and measure their degrees of anisotropy.
An isotropic tensor is represented by the origin of the diagram.
A uniaxial tensor is represented by three equivalent points on the uniaxial lines.
A biaxial tensor is represented by six equivalent points 
and each point belongs to one of the six domains divided by the uniaxial lines.
As shown in Fig.~3,
a generic point T in the diagram
is sandwiched between 
one of uniaxial half lines and one of maximally biaxial half lines,
which are denoted as $ U_j $ and $ B_{k} $.
The {\it degree of anisotropy} is measured by the radius OT $ = \alpha $ defined at (\ref{alpha}).
The angle between $ U_j $ and $ B_{k} $ is $ ( \pi /6 ) $ rad.
Let $ \theta $ be the magnitude of the angle formed by the half lines OT and $ U_j $
measured in radians.
Then we can define the {\it degree of biaxiality} $ \beta $ by
\begin{equation}
	\beta = \frac{6}{\pi}  \, \theta.
	\label{beta}
\end{equation}
Then $ \beta $ takes its value in the range $ 0 \le \beta \le 1 $.
Bates and Luckhurst~\cite{Bates2005}
gave another definition of biaxiality index $ \eta $,
which they called the {\it relative biaxiality},
\begin{equation}
	\eta 
	= \frac{ \lambda_1 - \lambda_2 }{ \lambda_3 }
\end{equation}
for $ \lambda_3 \ge 0 \ge \lambda_1 \ge \lambda_2 $.
It is equal to
\begin{equation}
	\eta 
	= \sqrt{3} \; \frac{ \varepsilon_2 }{ \varepsilon_1 }
	= \sqrt{3} \, \tan \theta
	= \frac{\tan \theta}{\tan ( \pi/6) }.
\end{equation}
The index $ \eta $ is a monotonically increasing function of $ \beta $
and it also takes its value in the range $ 0 \le \eta \le 1 $.
The intuitive meaning of $ \eta $ is also clear.
As another index of biaxiality we may define
\begin{equation}
	\zeta = \frac{ 2 \, \kappa }{ \alpha^6 } \, .
\end{equation}
The index $ \zeta $ also takes its value in $ 0 \le \zeta \le 1 $
since the value of $ \kappa $ is within
$ 0 \le \kappa \le \frac{1}{2} \alpha^6 $
as discussed at (\ref{kappa gamma}).

Here we need to mention that 
diagrams which are similar to our anisotropic diagram can be found in literatures.
Kralj, Virga, and {\v{Z}}umer~\cite{Kralj1999}
introduced a diagram which is equivalent to the anisotropic diagram.
A new point of our study is that
we use the diagram as a tool for comparing various tensorial quantities
and for measuring the degrees of anisotropies.
Another new point is that
we establish a relation 
between the microscopic quantity and the macroscopic quantity
in the anisotropic diagram as will be discussed in the next section.
For a comparison with Kralj's parameterization, 
Eq. (9) in their paper~\cite{Kralj1999},
we write the eigenvalues (\ref{co1})-(\ref{co3}) 
in terms of the variables (\ref{radius-angle}) as
\begin{equation}
	\lambda_1 
	= \sqrt{\frac{2}{3}} \, \alpha \, \cos \Big( \gamma - \frac{2 \pi}{3} \Big),
	\quad
	\lambda_2 
	= \sqrt{\frac{2}{3}} \, \alpha \, \cos \Big( \gamma + \frac{2 \pi}{3} \Big),
	\quad
	\lambda_3 
	= \sqrt{\frac{2}{3}} \, \alpha \, \cos  \gamma.
\end{equation}

\section{Micro-macro relation}\label{sect5}
In the previous section we introduced 
the anisotropy diagram, 
the degree of anisotropy $ \alpha $,
and the degree of biaxiality $ \beta $.
We have introduced also the geometric order parameter $ G $
which connects the microscopic tensorial quantity $ \vect{t} $
with the macroscopic tensorial quantity 
$ \bra \tilde{\vect{t}} \ket = G \vect{t} $.
In this section
we will discuss how 
the degrees of anisotropy of the macroscopic quantity is related to
the degrees of anisotropy of the microscopic quantity.
We will establish a diagrammatic relation between them,
which we call the micro-macro relation.

The idea of the micro-macro relation was inspired by
Ojima's idea, {\it Micro-macro duality}~\cite{Ojima2005}.
Micro-macro duality means bi-directional relations
between the microscopic quantum world and the macroscopic classical world.
Although our present consideration is restricted within classical physics,
the relation between the molecular quantity and the macroscopic quantity
can be regarded as an example of Micro-macro duality.

Let us confirm notation to be used below.
The orientation of a molecule in liquid crystal is described
by a three-dimensional rotation matrix $ R \in SO(3) $.
The molecules have various orientations 
and their statistical distribution is described by the probability density $ p(R) $.
Assume that each molecule has a physical quantity which is represented 
by a traceless symmetric tensor $ \vect{t} $.
When a molecule is turned by a matrix $ R $, the tensor is transformed to 
$ ( R \otimes R ) \vect{t} = R \, \vect{t} \, R^T $.
Then the average
\begin{equation}
	\bra \tilde{\vect{t}} \ket 
	= \int ( R \otimes R ) \vect{t} \, p(R) \, dR
	= \int ( R \, \vect{t} \, R^T ) \, p(R) \, dR
	= G \vect{t}
	\label{average}
\end{equation}
\begin{figure}[bt]
\begin{center}
\scalebox{0.55}{
\includegraphics{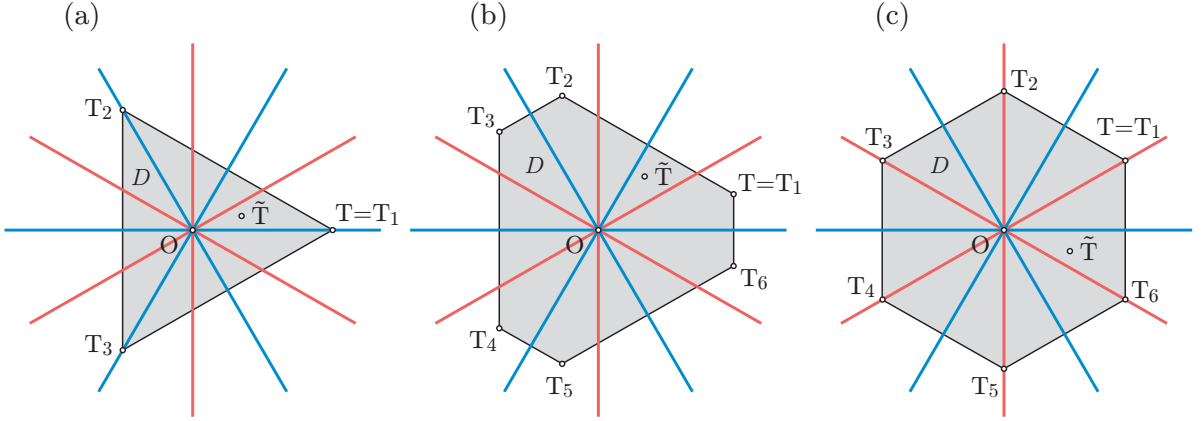}
}
\end{center}
\vspace{-2mm}
\caption{\label{fig4}
The points $ \{ {\rm T}_1, \cdots, {\rm T}_6 \} $ are equivalent points
representing a tensor $ \vect{t} $.
The point $ \tilde{\rm T} $ represents the average tensor 
$ \bra \tilde{\vect{t}} \ket $.
The point $ \tilde{\rm T} $ is always in the polygon $ D $ whose vertices are
$ \{ {\rm T}_1, \cdots, {\rm T}_6 \} $.
(a) T is uniaxial.
(b) T is generic.
(c) T is maximally biaxial.
}
\end{figure}
is the quantity observed at the macroscopic scale.
The tensor $ R \, \vect{t} \, {R}^T $ has the same set of eigenvalues as
$ \vect{t} $ for any rotation matrix $ R $.
However, in general, the principal axes of $ {R'} \, \vect{t} \, {R'}^T $
do not coincide with those of $ R \, \vect{t} \, R^T $
for different matrices $ R $ and $ R' $.
In other words, the matrices
$ \{ R \, \vect{t} \, R^T \} $ 
defined with various $ R \in SO(3) $ are not simultaneously diagonalizable.
Hence, it seems nontrivial to find a general relation which holds
between the eigenvalues of $ \vect{t} $ and those of $ \bra \tilde{\vect{t}} \ket $.
This is actually what we found and is called the micro-macro relation.

Let $ \{ \rm{T}_1, \cdots, \rm{T}_6 \} $ be the set of 
equivalent points in the anisotropy diagram
which corresponds to the tensor $ \vect{t} $.
The degrees of anisotropy and biaxiality of $ \vect{t} $ are written as
$ \alpha ( \vect{t} ) $ and $ \beta ( \vect{t} ) $, respectively.
Let $ \tilde{\rm{T}} $ be a point in the anisotropy diagram Fig.~4
which corresponds to $ \bra \tilde{\vect{t}} \ket $.
Let $ D $ be a convex polygon which has the points $ \{ \rm{T}_1, \cdots, \rm{T}_6 \} $
as its vertices.
Both the perimeter and the inner domain of $ D $ are included in $ D $.
Let P be an arbitrary point of $ D $.
Then the following two theorems hold.
\\
{\bf Theorem 1.} {\it
For any probability distribution $ p(R) $,
the average point $ \tilde{T} $ is in the polygon $ D $.
}
\\
{\bf Theorem 2.} {\it
For any point P in $ D $,
there exists a probability distribution $ p(R) $ such that
the average point $ \tilde{T} $ coincides with the point P.
}

Before showing proofs of these theorems,
we introduce three kinds of indices which enable us to compare
anisotropies of microscopic and macroscopic quantities.
We call the ratio of the degrees of anisotropy
\begin{equation}
	\chi_1 = \frac{ \alpha ( \bra \tilde{\vect{t}} \ket ) }{ \alpha (\vect{t}) }
\end{equation}
the {\it strength of realization of anisotropy}.
Theorem 1 implies that the length of the segment $ {\rm O} \tilde{\rm T} $
cannot be longer than the length of OT.
By definition, $ {\rm O} \tilde{\rm T} = \alpha ( \bra \tilde{\vect{t}} \ket ) $ and
OT $ = \alpha (\vect{t}) $.
Hence, their ratio $ \chi_1 $ is in the range $ 0 \le \chi_1 \le 1 $.
In particular, for the pair of
the microscopic singular tensor $ \vect{t}_\alpha $ and 
the macroscopic singular tensor $ \vect{u}_\alpha $ which satisfies
Eq.~(\ref{singular}),
the strength of realization of anisotropy $ \chi_1 $
is equal to the degree of order $ \sigma_\alpha $.

We call the ratio of the degrees of biaxiality
\begin{equation}
	\chi_2 = \frac{ \beta ( \bra \tilde{\vect{t}} \ket ) }{ \beta (\vect{t}) }
\end{equation}
the {\it strength of realization of biaxiality}.
The value of $ \chi_2 $ can be larger than 1.
In such a case it is said that 
{\it biaxiality is enhanced in the macroscopic phase}.
It can happen that 
$ \beta ( \bra \tilde{\vect{t}} \ket ) \ne 0 $
even when $ \beta (\vect{t}) = 0 $.
In such a case 
we formally write $ \chi_2 = \infty $ and say that 
{\it biaxiality is generated from uniaxial molecules}.
Conversely,
it also can happen that 
$ \beta ( \bra \tilde{\vect{t}} \ket ) = 0 $
although $ \beta (\vect{t}) \ne 0 $.
In such a case we have $ \chi_2 = 0 $ and say that 
{\it biaxiality is lost}
or
{\it uniaxiality is realized from biaxial molecules
in the macroscopic phase}.

We define the third index
\begin{equation}
	\chi_3 = 
	\mbox{sgn} \, 
	( \det \bra \tilde{\vect{t}} \ket \, \det \vect{t} ),
\end{equation}
and call it the {\it relative signature}.
Here sgn $ (x) $ denotes the signature of $ x $
and it takes its value in $ \{ 1, 0, -1 \} $.
When $ \chi_3 = 1 $, we say that the macroscopic phase is
{\it positively oriented} or it has {\it prolate order}.
When $ \chi_3 = -1 $, we say that the macroscopic phase is
{\it negatively oriented} or it has {\it oblate order}.

We would like to show another theorem.
Theorem 1 is a corollary of this theorem:
\\
{\bf Theorem 3.} {\it
Let $ \lambda_{\tiny \rm max} $ and $ \lambda_{\tiny \rm min} $
be the maximum and the minimum, respectively,
among the eigenvalues $ \{ \lambda_1, \lambda_2, \lambda_3 \} $ of $ \vect{t} $.
Let $ \{ \mu_1, \mu_2, \mu_3 \} $
be the eigenvalues of $ \bra \tilde{\vect{t}} \ket $.
Then, it holds that
}
\begin{equation}
	\lambda_{\tiny \rm min} \le \mu_r \le \lambda_{\tiny \rm max},
	\qquad
	r = 1,2,3.
	\label{theorem 3}
\end{equation}

{\it Proof of theorem 3: }
Let $ \vect{v}_r $ be a normalized eigenvector satisfying
$ \bra \tilde{\vect{t}} \ket \vect{v}_r = \mu_r \vect{v}_r $.
This equation is to be read as a multiplication of the matrix
$ \bra \tilde{\vect{t}} \ket $ on the vector $ \vect{v}_r $ as
\begin{equation}
	\sum_{j=1}^3 \:
	\bra \tilde{t}_{ij} \ket (v_r)_j = \mu_r (v_r)_i.
\end{equation}
We write an inner product of vectors $ \vect{v} $ and $ \vect{w} $
as $ \bra \vect{v}, \vect{w} \ket = \sum_{i=1}^3 v_i \, w_i $. Then we have
\begin{eqnarray}
	\mu_r
&=&	\mu_r \bra \vect{v}_r, \vect{v}_r \ket
	\nonumber \\
&=&	\bra \vect{v}_r, \bra \tilde{\vect{t}} \ket \vect{v}_r \ket
	\nonumber \\
&=&	\Big\langle 
	\vect{v}_r, \int ( R \, \vect{t} \, R^T ) \, p(R) \, dR \, \vect{v}_r 
	\Big\rangle
	\nonumber \\
&=&	\int \bra R^T \vect{v}_r, \vect{t} \, R^T \vect{v}_r \ket \, p(R) \, dR
	\nonumber \\
&=&	\int \tr 
	\Big\{ \vect{t} \, ( R^T \vect{v}_r \otimes R^T \vect{v}_r ) \Big\} \, p(R) \, dR
	\nonumber \\
&=&	\tr 
	\Big\{ 
	\vect{t} \int ( R^T \vect{v}_r \otimes R^T \vect{v}_r ) \, p(R) \, dR 
	\Big\}
	\nonumber \\
&=&	\tr ( \vect{t} \, \rho_r ).
	\label{mu}
\end{eqnarray}
In the last line we introduced the three-dimensional matrix $ \rho_r $
which is defined by
\begin{equation}
	\rho_r 
	= \int (R^T \vect{v}_r \otimes R^T \vect{v}_r ) \, p(R) \, dR 
	= \int R^T ( \vect{v}_r \otimes \vect{v}_r ) R \, p(R) \, dR.
\end{equation}
The matrix $ \rho_r $ is symmetric, non-negative and satisfies $ \tr \, \rho_r = 1 $.
On the other hand, let
\begin{equation}
	\vect{t} = \lambda_1 \Pi_1 + \lambda_2 \Pi_2 + \lambda_3 \Pi_3
\end{equation}
be the spectral decomposition of $ \vect{t} $.
The three-dimensional matrices $ \{ \Pi_q \} $ satisfy
$ (\Pi_q)^T = \Pi_q $, 
$ \Pi_q \Pi_s = \delta_{qs} \Pi_s $, 
$ \sum_{q=1}^3 \Pi_q = I $,
$ \tr \, \Pi_q = 1 $.
Substituting this into (\ref{mu}) we obtain
\begin{equation}
	\mu_r 
=	\tr ( \rho_r \, \vect{t} )
=	\sum_{q=1}^3 \tr( \rho_r \Pi_q ) \, \lambda_q
=	\sum_{q=1}^3 P_{rq} \, \lambda_q
\end{equation}
where $ P_{rq} = \tr( \rho_r \Pi_q ) $ are non-negative real numbers
and satisfy $ \sum_{q=1}^3 P_{rq} = 1 $. Hence,
\begin{eqnarray}
&&	\mu_r 
	=
	\sum_{q=1}^3 P_{rq} \, \lambda_q
	\; \le \;
	\sum_{q=1}^3 P_{rq} \, \lambda_{\tiny \rm max}
	=
	\lambda_{\tiny \rm max},
	\\ 
&&	\mu_r 
	=
	\sum_{q=1}^3 P_{rq} \, \lambda_q
	\; \ge \;
	\sum_{q=1}^3 P_{rq} \, \lambda_{\tiny \rm min}
	=
	\lambda_{\tiny \rm min}.
\end{eqnarray}
This ends the proof of theorem 3.
It is also interesting to note that $ \sum_{r=1}^3 \rho_r = I $ and therefore
$ \sum_{r=1}^3 P_{rq} = 1 $.

\begin{figure}[bt]
\begin{center}
\scalebox{0.55}{
\includegraphics{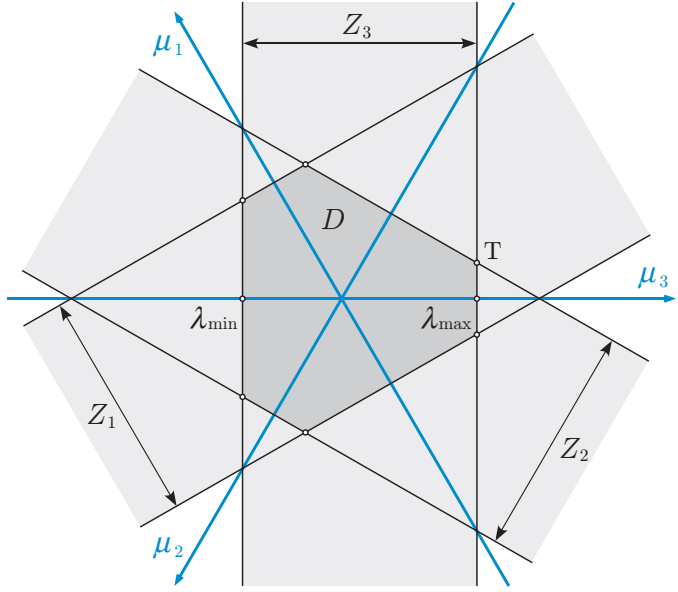}
}
\end{center}
\vspace{-2mm}
\caption{\label{fig5}
The intersection of the shaded bands, $ Z_1, Z_2, Z_3 $ defines the polygon $ D $.
The average point is restricted in $ D $.
}
\end{figure}
{\it Proof of theorem 1: }
{}From the construction of the anisotropy diagram
it is obvious that the set of points satisfying the inequality (\ref{theorem 3})
is the polygon $ D $.
See the Fig.~5.
This fact can be also verified via explicit calculations. 
The coordinate $ ( \tilde{\varepsilon}_1, \tilde{\varepsilon}_2 ) $
of the point $ \tilde{\rm{T}} $ is defined by
\begin{equation}
	\tilde{\varepsilon}_1 = \sqrt{\frac{3}{2}} \, \mu_3,
	\qquad
	\tilde{\varepsilon}_2 
	= \frac{1}{\sqrt{2}} ( \mu_1 - \mu_2 ).
	\label{tildevare}
\end{equation}
They satisfy the equation similar to (\ref{co1})-(\ref{co3}),
\begin{eqnarray}
&&	\mu_1 
	= - \frac{1}{\sqrt{6}} \, \tilde{\varepsilon}_1 
	+ \frac{1}{\sqrt{2}} \, \tilde{\varepsilon}_2
	= \sqrt{\frac{2}{3}} \cdot \frac{1}{2}
	( -1,   \sqrt{3} ) 
	\begin{pmatrix} \tilde{\varepsilon}_1 \\ \tilde{\varepsilon}_2 \end{pmatrix},
	\label{tilco1}
	\\
&&	\mu_2 
	= - \frac{1}{\sqrt{6}} \, \tilde{\varepsilon}_1 
	- \frac{1}{\sqrt{2}} \, \tilde{\varepsilon}_2
	= \sqrt{\frac{2}{3}} \cdot \frac{1}{2}
	( -1, - \sqrt{3} ) 
	\begin{pmatrix} \tilde{\varepsilon}_1 \\ \tilde{\varepsilon}_2 \end{pmatrix},
	\label{tilco2}
	\\
&&	\mu_3 
	= \sqrt{\frac{2}{3}} \, \tilde{\varepsilon}_1
	= \sqrt{\frac{2}{3}} \, (1,0) 
	\begin{pmatrix} \tilde{\varepsilon}_1 \\ \tilde{\varepsilon}_2 \end{pmatrix}.
	\label{tilco3}
\end{eqnarray}
Hence the sets of points restricted by the inequality (\ref{theorem 3}),
\begin{eqnarray}
&&	Z_1 = 
	\{ ( \tilde{\varepsilon}_1, \tilde{\varepsilon}_2 ) \, | \,
	\lambda_{\tiny \rm min} \le 
	- \frac{1}{\sqrt{6}} \, \tilde{\varepsilon}_1 
	+ \frac{1}{\sqrt{2}} \, \tilde{\varepsilon}_2
	\le \lambda_{\tiny \rm max} \},
\\
&&	Z_2 = 
	\{ ( \tilde{\varepsilon}_1, \tilde{\varepsilon}_2 ) \, | \,
	\lambda_{\tiny \rm min} \le 
	- \frac{1}{\sqrt{6}} \, \tilde{\varepsilon}_1 
	- \frac{1}{\sqrt{2}} \, \tilde{\varepsilon}_2
	\le \lambda_{\tiny \rm max} \},
\\
&&	Z_3 = 
	\{ ( \tilde{\varepsilon}_1, \tilde{\varepsilon}_2 ) \, | \,
	\lambda_{\tiny \rm min} \le 
	\sqrt{\frac{2}{3}} \, \tilde{\varepsilon}_1
	\le \lambda_{\tiny \rm max} \},
\end{eqnarray}
are drawn as three shaded bands in Fig.~5.
Their intersection $ Z_1 \cap Z_2 \cap Z_3 $ is nothing but the polygon $ D $.
This observation proves theorem 1.

{\it Proof of theorem 2:} 
First, let us note the following simple fact.
Suppose that two traceless symmetric tensors $ \vect{t}_0 $ and $ \vect{t}_1 $ 
are simultaneously diagonalizable.
In other words, they have common principal axes.
Let $ \rm{T}_0 $ and $ \rm{T}_1 $ be their representing points in the anisotropy diagram.
Then the weighted sum
\begin{equation}
	\vect{t}_w = (1-w) \vect{t}_0 + w \vect{t}_1 
\end{equation}
with a real number $ w $ $ ( 0 \le w \le 1 ) $ 
is also a traceless symmetric tensor
and diagonalizable simultaneously with $ \vect{t}_0 $ and $ \vect{t}_1 $.
It is easily verified that
the point $ {\rm T}_w $ representing $ \vect{t}_w $ in the anisotropy diagram
divides the segment  $ \rm{T}_0 \rm{T}_1 $ in the ratio $ w : (1-w) $.

Second, let us note that any point P of a convex polygon can be expressed 
as a weighted sum of the vertices of the polygon.
We can chose real numbers $ w_1, \cdots, w_6 $ such that
\begin{equation}
	\vect{t}_{\rm P} = \sum_{i=1}^6 w_i \vect{t}_i,
	\qquad
	\sum_{i=1}^6 w_i = 1,
	\qquad
	0 \le w_i \le 1
\end{equation}
\begin{figure}[bt]
\begin{center}
\scalebox{0.55}{
\includegraphics{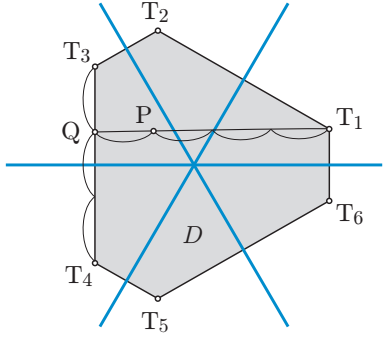}
}
\end{center}
\vspace{-2mm}
\caption{\label{fig6}
The point Q divides the edge $ {\rm T}_3 {\rm T}_4 $ into the ratio $ 1:2 $.
The point P divides the segment $ {\rm T}_1 {\rm Q} $ into the ratio $ 3:1 $.
Thus,
Q $ = \frac{2}{3} {\rm T}_3 + \frac{1}{3} {\rm T}_4$,
P $ = \frac{1}{4} {\rm T}_1 + \frac{3}{4} {\rm Q} 
= \frac{1}{4} {\rm T}_1 + \frac{1}{2} {\rm T}_3 + \frac{1}{4} {\rm T}_4 $.
Every point in the polygon $ D $ can be expressed as a weight sum of the vertices.
}
\end{figure}
For example, the point Q in Fig.~6 is given by
\begin{equation}
	\vect{t}_{\rm Q} = \frac{2}{3} \vect{t}_3 + \frac{1}{3} \vect{t}_4
\end{equation}
and the point P is given by
\begin{equation}
	\vect{t}_{\rm P} 
	= \frac{1}{4} \vect{t}_1 + \frac{3}{4} \vect{t}_{\rm Q}
	= \frac{1}{4} \vect{t}_1 + \frac{1}{2} \vect{t}_3 + \frac{1}{4} \vect{t}_4.
\end{equation}
For a given point P the set of weights $ ( w_1, \cdots, w_6 ) $ is not unique
but uniqueness is not necessary.

Third, remember that the vertices of the polygon $ D $ in the anisotropy diagram
are related to each other by reflections $ \phi_1, \phi_2, \phi_3 $.
Note that these reflections in the diagram can be generated by rotations in the real space.
If we define
\begin{equation}
	K_1 =
	\begin{pmatrix}
	1 & 0 & 0 \\
	0 & 0 &-1 \\
	0 & 1 & 0 
	\end{pmatrix}, 
	\qquad
	K_2 =
	\begin{pmatrix}
	0 & 0 & 1 \\
	0 & 1 & 0 \\
	-1& 0 & 0 
	\end{pmatrix}, 
	\qquad
	K_3 =
	\begin{pmatrix}
	0 &-1 & 0 \\
	1 & 0 & 0 \\
	0 & 0 & 1 
	\end{pmatrix}, 
	\label{Ks}
\end{equation}
then the reflection mapping $ \phi_i $ introduced in Sec.~\ref{sect4}
is equivalent to the rotation $ \phi_i \vect{t} = K_i \, \vect{t} \, K_i^T $.
Hence, for the vertices 
$ \vect{t}_1 ( = \vect{t}), \vect{t}_2, \cdots, \vect{t}_6 $
of the polygon there exists a set of rotation matrices
$ R_1, $ $ R_2, \cdots, $ $ R_6 $ such that $ \vect{t}_i = R_i \, \vect{t} \, R_i^T $.

Finally, combining the above arguments we obtain
\begin{equation}
	\vect{t}_{\rm P} 
	= \sum_{i=1}^6 w_i \vect{t}_i
	= \sum_{i=1}^6 w_i ( R_i \, \vect{t} \, R_i^T ),
\end{equation}
which should be compared with Eq.~(\ref{average}).
This means that the set of weights $ ( w_1, \cdots, w_6 ) $ is
a probability distribution
which yields the point P as the average.
This proves theorem 2.

The collection of theorem 1, 2, and 3 is called the {\it micro-macro relation}.

\section{Examples}\label{sect6}
In this section we will demonstrate calculations of the macroscopic tensors
by assuming simple probability distributions.
In the first example we will show that
uniaxial molecules can generate a biaxial order at the macroscopic scale.
In the second example we will show that
uniaxial molecules can exhibit a negatively oriented uniaxial phase.
The third example has a continuous probability distribution and
will exhibit the same result as the second one.
In the fourth example we will show that
biaxial molecules can exhibit a uniaxial order.

In the first example we assume that the molecule has a physical quantity
\begin{equation}
	\vect{t} = \frac{1}{ \sqrt{6} }
	\begin{pmatrix}
	-1& 0 & 0 \\
	0 &-1 & 0 \\
	0 & 0 & 2
	\end{pmatrix},
\end{equation}
which has positive uniaxiality on the $ z $-axis.
The anisotropy diagram for this is shown in Fig.~7 (a).
The coordinate of the representing point T is
$ ( \varepsilon_1, \varepsilon_2 ) = (1,0) $.
Assume that 2/3 of molecules are aligned in the $ z $-direction
and 1/3 of molecules turn into the $ x $-direction.
Then the average of the tensorial quantity is
\begin{eqnarray}
	\bra \tilde{\vect{t}} \ket_1
	= 
	\frac{2}{3} \vect{t} +
	\frac{1}{3} ( K_2 \vect{t} K_2^T )
&=&
	\frac{2}{3} \cdot \frac{1}{ \sqrt{6} }
	\begin{pmatrix}
	-1& 0 & 0 \\
	0 &-1 & 0 \\
	0 & 0 & 2
	\end{pmatrix}
	+ \frac{1}{3} \cdot \frac{1}{ \sqrt{6} }
	\begin{pmatrix}
	2 & 0 & 0 \\
	0 &-1 & 0 \\
	0 & 0 &-1
	\end{pmatrix}
\nonumber \\
&=& 
	\frac{1}{ \sqrt{6}}
	\begin{pmatrix}
	0 & 0 & 0 \\
	0 &-1 & 0 \\
	0 & 0 & 1
	\end{pmatrix}.
\end{eqnarray}
The coordinate of the point $ \tilde{\rm T}_1 $ 
corresponding to $ \bra \tilde{\vect{t}} \ket_1 $ is
$ ( \tilde{\varepsilon}_1, \tilde{\varepsilon}_2 ) 
= \frac{1}{2} ( 1, \frac{1}{\sqrt{3}} ) $
and it lies on the half line $ B_{-y,+z} $.
The maximal biaxiality is realized in this case.
The strength of realization of anisotropy is
\begin{equation}
	\chi_1 = 
	\frac{ \alpha ( \bra \tilde{\vect{t}} \ket_1 ) }{ \alpha (\vect{t}) }
	= \sqrt{\frac{2}{6}}
	= \frac{1}{\sqrt{3}}.
\end{equation}
The degrees of biaxiality are
$ \beta (\vect{t}) = 0 $ and
$ \beta (\bra \tilde{\vect{t}} \ket_1 ) = 1 $.
Hence the strength of realization of biaxiality is
$ \chi_2 = \infty $.

Next, assume that
 1/2 of molecules turn in the $ x $-direction
and 1/2 of molecules turn into the $ y $-direction.
Then the average is
\begin{eqnarray}
	\bra \tilde{\vect{t}} \ket_2
	= 
	\frac{1}{2} ( K_2 \vect{t} K_2^T ) +
	\frac{1}{2} ( K_1 \vect{t} K_1^T )
&=&
	\frac{1}{2} \cdot \frac{1}{ \sqrt{6} }
	\begin{pmatrix}
	2 & 0 & 0 \\
	0 &-1 & 0 \\
	0 & 0 &-1
	\end{pmatrix}
	+ \frac{1}{2} \cdot \frac{1}{ \sqrt{6} }
	\begin{pmatrix}
	-1& 0 & 0 \\
	0 & 2 & 0 \\
	0 & 0 &-1
	\end{pmatrix}
\nonumber \\
	&=& 
	\frac{1}{ 2 \sqrt{6} }
	\begin{pmatrix}
	1 & 0 & 0 \\
	0 & 1 & 0 \\
	0 & 0 &-2
	\end{pmatrix}.
	\label{example t2}
\end{eqnarray}
\begin{figure}[bt]
\begin{center}
\scalebox{0.55}{
\includegraphics{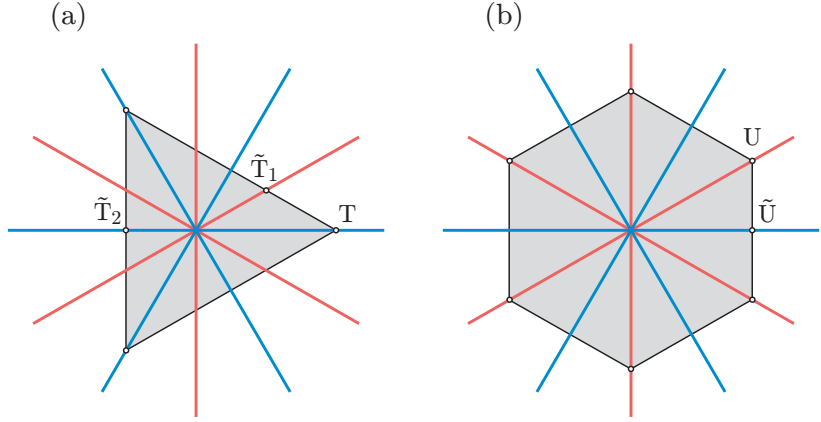}
}
\end{center}
\vspace{-2mm}
\caption{\label{fig7}
Examples of calculation of the average tensor $ \tilde{\rm T} $
from the microscopic tensor T.
(a) The uniaxial tensor T can generate the biaxial $ \tilde{\rm T}_1 $.
It can also generate the uniaxial $ \tilde{\rm T}_2 $ with the inverted signature.
(b) The biaxial tensor U can generate the uniaxial $ \tilde{\rm U} $.
}
\end{figure}
The coordinate of the point $ \tilde{\rm T}_2 $ in Fig.~7 (a)
corresponding to $ \bra \tilde{\vect{t}} \ket_2 $ is
$ ( \tilde{\varepsilon}_1, \tilde{\varepsilon}_2 ) = (- \frac{1}{2}, 0 ) $.
The strength of realization of anisotropy is
$ \chi_1 = \frac{1}{2} $.
The degrees of biaxiality remains
$ \beta (\bra \tilde{\vect{t}} \ket_2 ) = 0 $.
The relative signature is $ \chi_3 = -1 $ in this case.
Thus the macroscopic phase has oblate order.

In the above two examples,
the probability distribution $ p(R) $ that we assumed had pointwise support,
namely,
the integral in Eq.~(\ref{average}) was replaced by summation.
Here we show an example which has a continuous probability distribution.
We define a rotation matrix
\begin{equation}
	K_3 ( \phi ) =
	\begin{pmatrix}
	\cos \phi & -\sin \phi & 0 \\
	\sin \phi &  \cos \phi & 0 \\
	0 & 0 & 1
	\end{pmatrix}
\end{equation}
which is parameterized by an angle $ \phi $.
Assume that the molecules are turned as
$ \vect{t} \mapsto K_3 ( \phi ) K_2 \, \vect{t} \, K_2^T K_3 ( \phi )^T $
with a probability distribution which is uniform with respect to 
the variable $ \phi $.
Then the average becomes
\begin{eqnarray}
	\bra \tilde{\vect{t}} \ket_3
	&=& 
	\frac{1}{2 \pi} \int_0^{2 \pi} 
	( K_3 ( \phi ) K_2 \, \vect{t} \, K_2^T K_3 ( \phi )^T )
	\, d \phi
\nonumber \\
	&=& 
	\frac{1}{2 \pi} \int_0^{2 \pi} 
	\frac{1}{ \sqrt{6} } \!
	\begin{pmatrix}
	2 \cos^2 \phi - \sin^2 \phi & 3 \cos \phi \sin \phi & 0 \\
	3 \cos \phi \sin \phi & 2 \sin^2 \phi - \cos^2 \phi & 0 \\
	0 & 0 & -1
	\end{pmatrix} \!
	d \phi
\nonumber \\
	&=& 
	\frac{1}{ \sqrt{6} }
	\begin{pmatrix}
	\frac{1}{2} & 0 & 0 \\
	0 & \frac{1}{2}  & 0 \\
	0 & 0 & -1
	\end{pmatrix}.
\end{eqnarray}
This result is the same as (\ref{example t2}).

In the fourth example we assume that the molecule has a physical quantity
\begin{equation}
	\vect{u} = \frac{1}{\sqrt{2}}
	\begin{pmatrix}
	0 & 0 & 0 \\
	0 &-1 & 0 \\
	0 & 0 & 1
	\end{pmatrix}.
\end{equation}
The representing point U is shown in Fig.~7 (b).
Its coordinate is
$ ( \varepsilon_1, \varepsilon_2 ) = \frac{1}{2} (\sqrt{3}, 1) $.
It lies on the half line $ B_{-y,+z} $
and has the maximum biaxiality $ \beta ( \vect{u} ) = 1 $.
Assume that 1/2 of molecules are aligned in the same orientation
and 1/2 of molecules are turned about the $ z $-axis by the right angle.
Then the average becomes
\begin{eqnarray}
	\bra \tilde{\vect{u}} \ket 
	= 
	\frac{1}{2} \vect{t} +
	\frac{1}{2} ( K_3 \vect{t} K_3^T )
&=&
	\frac{1}{2} \cdot \frac{1}{\sqrt{2}}
	\begin{pmatrix}
	0 & 0 & 0 \\
	0 &-1 & 0 \\
	0 & 0 & 1
	\end{pmatrix}
	+ \frac{1}{2} \cdot \frac{1}{\sqrt{2}}
	\begin{pmatrix}
	-1& 0 & 0 \\
	0 & 0 & 0 \\
	0 & 0 & 1
	\end{pmatrix}
\nonumber \\
&=&
	\frac{1}{2 \sqrt{2}}
	\begin{pmatrix}
	-1& 0 & 0 \\
	0 &-1 & 0 \\
	0 & 0 & 2
	\end{pmatrix},
\end{eqnarray}
which has positive uniaxiality on the $ z $-axis.
In this case biaxiality is lost in the macroscopic phase.
The coordinate of the representing point $ \tilde{\rm U} $ is
$ ( \tilde{\varepsilon}_1, \tilde{\varepsilon}_2 ) 
= \frac{\sqrt{3}}{2} ( 1, 0 ) $.
The strength of realization of anisotropy is
$ \chi_1 = \frac{\sqrt{3}}{2} $.

\section{Order parameters for $ D_{2h} $-symmetric molecules}\label{sect7}
Here we apply the method of the geometric order parameter
to platelet molecules.
By platelet molecules we mean molecules which possess
the point group $ D_{2h} $ as its symmetry.
The shape of a platelet molecule is invariant
under reflections on the
$ yz $-plane, $ xz $-plane and $ xy $-plane.
The $ D_{2h} $ symmetry is a group generated by
\begin{equation}
	\begin{pmatrix}
	-1& 0 & 0 \\
	0 & 1 & 0 \\
	0 & 0 & 1
	\end{pmatrix},
	\qquad
	\begin{pmatrix}
	1 & 0 & 0 \\
	0 &-1 & 0 \\
	0 & 0 & 1
	\end{pmatrix},
	\qquad
	\begin{pmatrix}
	1 & 0 & 0 \\
	0 & 1 & 0 \\
	0 & 0 &-1
	\end{pmatrix}.
	\label{D2h}
\end{equation}
A physical quantity $ \vect{t} $ of the platelet molecule must be
invariant under the actions of $ D_{2h} $.
Namely, it is required that 
$ R \vect{t} R^T = \vect{t} $ for any $ R \in D_{2h} $.
This implies that the off-diagonal elements satisfy
$ - t_{ij} = t_{ij} $ for $ i \ne j $, hence $ t_{ij} = 0 $.
Thus, only the diagonal elements $ t_{ii} $ can be nonzero.
Moreover, we assume that 
the associated macroscopic quantity 
$ \bra \tilde{\vect{t}}_{ab} \ket = \sum_{i,j} G_{abij} t_{ij} $ 
is also invariant under the actions of $ D_{2h} $.
Namely, it is assumed that 
$ R \bra \tilde{\vect{t}} \ket R^T = \bra \tilde{\vect{t}} \ket $
for any $ R \in D_{2h} $.
Then only the diagonal elements $ \bra \tilde{\vect{t}}_{aa} \ket $
can be nonzero.
The invariance under the $ D_{2h} $ transformations
requires that
\begin{equation}
	\sum_{c,d=1}^3 R_{ac} \, R_{bd} \, G_{cdij} = G_{abij},
	\qquad
	\sum_{i,j=1}^3 G_{abij} \, R_{ik} \, R_{jl} = G_{abkl}
	\qquad \mbox{for} \quad
	( R_{ij} ) \in D_{2h}.
\end{equation}
These requirements for the geometric order parameter $ G $ are equivalent to
\begin{equation}
	- G_{abij} = G_{abij} \quad \mbox{for} \quad a \ne b,
	\qquad
	- G_{abij} = G_{abij} \quad \mbox{for} \quad i \ne j
\end{equation}
Therefore,
\begin{equation}
	G_{abij} = 0 \quad \mbox{for} \quad a \ne b,
	\qquad
	G_{abij} = 0 \quad \mbox{for} \quad i \ne j.
\end{equation}
Hence, elements which can be nonzero are $ G_{aaii} $ with $ a,i =1,2,3 $.
In the following we abbreviate it as $ G_{aaii} = G_{a;i} $.
The nine components $ \{ G_{a;i} \} $ are imposed the traceless condition
\begin{equation}
	\sum_{a=1}^3 G_{a;i} = 0,
	\qquad
	\sum_{i=1}^3 G_{a;i} = 0.
	\label{traceless}
\end{equation}
Hence, only four components among $ \{ G_{a;i} \} $ are independent.

Dummur and Toriyama~\cite{Dunmur1999, Bates2005} defined four parameters $ S, D, P, C $ as
\begin{eqnarray}
&&	S = S^{ZZ}_{zz} = \frac{3}{2} G_{3;3},
	\label{S}
	\\
&&	D = S^{ZZ}_{xx} - S^{ZZ}_{yy} = \frac{3}{2} ( G_{3;1} - G_{3;2} ),
	\label{D}
	\\
&&	P = S^{XX}_{zz} - S^{YY}_{zz} = \frac{3}{2} ( G_{1;3} - G_{2;3} ),
	\label{P}
	\\
&&	C 
	= ( S^{XX}_{xx} - S^{YY}_{xx} ) - ( S^{XX}_{yy} - S^{YY}_{yy} )
	= \frac{3}{2} ( G_{1;1} - G_{2;1} - G_{1;2} + G_{2;2} ).
	\label{CC}
\end{eqnarray}
These are almost equal to the parameters $ S, T, U, V $ 
which Straley~\cite{Straley1974} introduced.
Only the differences between $ (S,D,P,C) $ and $ (S,T,U,V) $ are multiplicative factors
as explained in the reference~\cite{Bates2005}.
Here $ S^{ab}_{ij} $ are elements of the de Gennes ordering matrix (\ref{de Gennes}).
The indices $ a,b= X,Y,Z $ specify axes of the laboratory frame
while $ i,j= x,y,z $ specify axes of the molecular frame.
Physical meanings of these parameters are explained as follows.
The parameter $ S $ is an index to measure
how strongly
the uniaxiality of the molecule manifests itself 
as the uniaxiality of the macroscopic phase.
$ D $ measures
how strongly
the biaxiality of the molecule manifests itself 
as the uniaxiality of the macroscopic phase.
$ P $ is an index of
the strength with which
the molecular uniaxiality generates the macroscopic biaxiality.
$ C $ represents
how strongly
the molecular biaxiality generates the macroscopic biaxiality.

The parameters $ (S,D,P,C) $ can be written in various forms.
In the present case the microscopic quantity $ \vect{t} $ 
and the macroscopic quantity $ \bra \tilde{\vect{t}} \ket $
can be diagonalized as
$ \vect{t} = \mbox{diag} ( \lambda_1, \lambda_2, \lambda_3 ) $ and
$ \bra \tilde{\vect{t}} \ket = \mbox{diag} ( \mu_1, \mu_2, \mu_3 ) $.
Then the relation $ \bra \tilde{\vect{t}} \ket = G \vect{t} $ is written as
\begin{equation}
	\begin{pmatrix}
	\mu_1 \\ \mu_2 \\ \mu_3 
	\end{pmatrix}
	=
	\begin{pmatrix}
	G_{1;1} & G_{1;2} & G_{1;3} \\
	G_{2;1} & G_{2;2} & G_{2;3} \\
	G_{3;1} & G_{3;2} & G_{3;3} 
	\end{pmatrix}
	\begin{pmatrix}
	\lambda_1 \\ \lambda_2 \\ \lambda_3 
	\end{pmatrix}.
	\label{mu-lambda}
\end{equation}
The set of equations (\ref{traceless}), (\ref{S})-(\ref{CC}) is solved
for the geometric order parameters as
\begin{equation}
	\begin{pmatrix}
	G_{1;1} & G_{1;2} & G_{1;3} \\
	G_{2;1} & G_{2;2} & G_{2;3} \\
	G_{3;1} & G_{3;2} & G_{3;3} 
	\end{pmatrix}
	= \frac{1}{6}
	\begin{pmatrix}
	S-D-P+C &
	S+D-P-C &
	-2S +2P \\
	S-D+P-C &
	S+D+P+C &
	-2S -2P \\
	-2S +2D &
	-2S -2D &
	4S
	\end{pmatrix}.
	\label{solved}
\end{equation}
Eqs.~(\ref{mu-lambda}) and (\ref{solved}) yield
\begin{eqnarray}
&&	\mu_3 
	= \frac{1}{3} (-S+D) \lambda_1
	+ \frac{1}{3} (-S-D) \lambda_2
	+ \frac{2}{3} S \lambda_3
	= \frac{1}{3} D ( \lambda_1 - \lambda_2 )
	+ S \lambda_3,
	\\
&&	\mu_1 - \mu_2 
	= \frac{1}{3} (-P+C) \lambda_1
	+ \frac{1}{3} (-P-C) \lambda_2
	+ \frac{2}{3} P \lambda_3
	= \frac{1}{3} C ( \lambda_1 - \lambda_2 )
	+ P \lambda_3,
\end{eqnarray}
where we used $ \lambda_1 + \lambda_2 + \lambda_3 = 0 $.
They can be put in the form
\begin{equation}
	\begin{pmatrix}
	\mu_3 \\ \mu_1 - \mu_2 
	\end{pmatrix}
	=
	\begin{pmatrix}
	S & \frac{1}{3} D \\
	P & \frac{1}{3} C
	\end{pmatrix}
	\begin{pmatrix}
	\lambda_3 \\ \lambda_1 - \lambda_2
	\end{pmatrix}.
\end{equation}
This equation is consistent with the interpretation 
of the parameters $ (S,D,P,C) $ explained above.
It can be expressed in terms of the anisotropy coordinates
(\ref{varepsilon_1}), (\ref{varepsilon_2}), (\ref{tildevare}) as
\begin{equation}
	\begin{pmatrix}
	\tilde{\varepsilon}_1 \\ \tilde{\varepsilon}_2
	\end{pmatrix}
	=
	\begin{pmatrix}
	S                    & \frac{1}{\sqrt{3}} D \\
	\frac{1}{\sqrt{3}} P & \frac{1}{3} C
	\end{pmatrix}
	\begin{pmatrix}
	\varepsilon_1 \\ \varepsilon_2
	\end{pmatrix}
	=
	\begin{pmatrix}
	\hat{G}_{55} &
	\hat{G}_{54} \\
	\hat{G}_{45} &
	\hat{G}_{44} 
	\end{pmatrix}
	\begin{pmatrix}
	\varepsilon_1 \\ \varepsilon_2
	\end{pmatrix}.
	\label{D2h geo para}
\end{equation}
On the other hand, the elements of the reduced ordering matrix $ \hat{G} $ 
are calculated from the definitions
(\ref{xis}), (\ref{reduced ordering matrix})
with the help of (\ref{traceless}) as
\begin{eqnarray*}
&&	\hat{G}_{55}
	= \frac{1}{6}
	( G_{1;1}+G_{1;2}-2G_{1;3}+G_{2;1}+G_{2;2}-2G_{2;3}-2G_{3;1}-2G_{3;2}+4G_{3;3} )
	= \frac{3}{2} G_{3;3} 
	= S,
	\\
&&	\hat{G}_{54}
	= \frac{1}{2 \sqrt{3}}
	(-G_{1;1}+G_{1;2}-G_{2;1}+G_{2;2}+2G_{3;1}-2G_{3;2} )
	= \frac{\sqrt{3}}{2}
	( G_{3;1}-G_{3;2} )
	= \frac{1}{\sqrt{3}} D,
	\\
&&	\hat{G}_{45}
	= \frac{1}{2 \sqrt{3}}
	(-G_{1;1}-G_{1;2}+2G_{1;3}+G_{2;1}+G_{2;2}-2G_{2;3} )
	= \frac{\sqrt{3}}{2}
	( G_{1;3}-G_{2;3} )
	= \frac{1}{\sqrt{3}} P,
	\\
&&	\hat{G}_{44}
	= \frac{1}{2}
	( G_{1;1}-G_{1;2}-G_{2;1}+G_{2;2} )
	= \frac{1}{3} C.
\end{eqnarray*}
This result is consistent with (\ref{D2h geo para}).

\section{Landau-de Gennes free energy}\label{sect8}
The Landau-de Gennes free energy is a standard tool
for analysis of phase structures of liquid crystals.
The Landau-de Gennes free energy is a polynomial function 
$ {\mathscr F} (\vect{a}) $
of a collection of macroscopic quantities, which is denoted as $ \vect{a} $.
The quantities $ \vect{a} $ play the role of order parameters, too.
The free energy should be invariant under spatial rotations of the variables.
This requirement is symbolically written as
$ {\mathscr F} ( R \vect{a}) = {\mathscr F} (\vect{a}) $.
The coefficients $ c_1, c_2, \cdots $ in the polynomial 
$ {\mathscr F} (\vect{a}) = c_1 \vect{a} + c_2 \vect{a}^2 + \cdots $
may depend on various external physical parameters 
like temperature or density.
It is required that in an equilibrium state
the free energy takes its minimum value.
Thus the values of the order parameters $ \vect{a} $ are determined 
as the minimizer of the free energy.
Then the symmetry of the equilibrium phase is determined
by the values of the order parameters,
which are also functions of the external parameters.
This is a usual routine to analyze the phase structure
using the Landau-de Gennes free energy.

In this section we will explain a general prescription
to formulate the Landau-de Gennes free energy 
of arbitrary shape molecules.
Here we mainly consider nematic phases, which are translationally invariant.
A possible generalization for smectic phases will be discussed briefly.
Later we will apply our prescription to
the $ D_{2h} $-symmetric molecules.

Assume that each molecule has
microscopic quantities $ \vect{a}_0 $, $ \vect{b}_0 $, $ \cdots $,
which are symmetric tensors.
They can be 
the dielectric susceptibility tensor or 
the magnetic susceptibility tensor of a molecule.
We can assume that they are traceless.
If $ \vect{a}_1 $ is not traceless,
we can take the traceless component by subtracting its trace to define
$ \vect{a}_0 = \vect{a}_1 - \frac{1}{3} \tr( \vect{a}_1 ) $.
At a macroscopic scale we measure
physical quantities, $ \vect{a} $, $ \vect{b}, \cdots $, 
which are ensemble averages of microscopic quantities.
The geometric order parameter $ G $ relates them as
$ G \vect{a}_0 = \vect{a} $.
The tensor $ \vect{a} $ is transformed  
under a rotation $ R \in SO(3) $ as $ \vect{a} \mapsto R \vect{a} R^T $.
A polynomial function
$ I ( \vect{a}, \vect{b}, \cdots )$ which satisfies
\begin{equation}
	I ( R \vect{a} R^T, R \vect{b} R^T, \cdots )
	=
	I ( \vect{a}, \vect{b}, \cdots )
\end{equation}
for arbitrary $ R \in SO(3) $ is called an {\it invariant polynomial}.
Then the Landau-de Gennes free energy is defined as a function of $ G $,
\begin{equation}
	{\mathscr F} ( G )
	=
	I ( G \vect{a}_0, G \vect{b}_0, \cdots ).
\end{equation}
The invariant polynomials up to the fourth order are listed as
\begin{eqnarray}
&&	I_1 
	= \tr ( \vect{a} \vect{b} ) 
	= \tr ( \vect{b} \vect{a} ),
	\label{I1}
	\\
&&	I_2 
	= \tr ( \vect{a} \vect{b} \vect{c} )
	= \tr ( \vect{a} \vect{c} \vect{b} ),
	\label{I2}
	\\
&&	I_3
	= \tr ( \vect{a} \vect{b} \vect{c} \vect{d} )
	= \tr ( \vect{d} \vect{c} \vect{b} \vect{a} ),
	\label{I3}
	\\
&&	I_4
	= \tr ( \vect{a} \vect{b} \vect{d} \vect{c} )
	= \tr ( \vect{c} \vect{d} \vect{b} \vect{a} ),
	\\
&&	I_5
	= \tr ( \vect{a} \vect{c} \vect{b} \vect{d} )
	= \tr ( \vect{d} \vect{b} \vect{c} \vect{a} ),
	\\
&&	I_6
	= \tr ( \vect{a} \vect{b} ) \, \tr ( \vect{c} \vect{d} ),
	\\
&&	I_7
	= \tr ( \vect{a} \vect{c} ) \, \tr ( \vect{b} \vect{d} ),
	\label{I7} 
	\\
&&	I_8
	= \tr ( \vect{a} \vect{d} ) \, \tr ( \vect{b} \vect{c} ).
	\label{I8}
\end{eqnarray}
The equal signs in the above equations hold since 
$ \vect{a}^T = \vect{a} $ and 
\begin{equation}
	  \tr ( \vect{a} \vect{b} \vect{c} ) 
	= \tr ( \vect{a} \vect{b} \vect{c} )^T 
	= \tr ( \vect{c}^T \vect{b}^T \vect{a}^T ) 
	= \tr ( \vect{c} \vect{b} \vect{a} )
	= \tr ( \vect{a} \vect{c} \vect{b} ).
\end{equation}
Other polynomials like
$ \tr ( \vect{a} ) \tr ( \vect{b} ) $,
$ \tr ( \vect{a} ) \tr ( \vect{b} \vect{c} ) $,
$ \tr ( \vect{a} ) \tr ( \vect{b} ) \, \tr ( \vect{c} \vect{d} ) $
become zero because $ \vect{a}, \vect{b}, \cdots $ are traceless.
Using the representation theory of the rotation group, we can prove the following theorem:
\\
{\bf Theorem 4.} {\it
Among the polynomials of traceless symmetric tensors,
there is only one linearly independent invariant of the second order,
that is $ I_1 $.
There is one linearly independent invariant of the third order, 
that is $ I_2 $.
There are five linearly independent invariants of the fourth order.
The six invariants $ \{ I_3, \cdots, I_8 \} $ always satisfy}
\begin{equation}
	I_3 + I_4 + I_5 - \frac{1}{2} ( I_6 + I_7 + I_8 ) = 0.
	\label{dependence}
\end{equation}
{\it
Hence, only five among $ \{ I_3, \cdots, I_8 \} $ are linearly independent.
}

A proof of this theorem is given in the appendix.
If there are four independent microscopic quantities
$ \vect{a}_0 $, $ \vect{b}_0 $, $ \vect{c}_0 $, $ \vect{d}_0 $,
the Landau-de Gennes free energy up to the fourth order is constructed 
by substituting 
$ \vect{a} = G \vect{a}_0 $, 
$ \vect{b} = G \vect{b}_0 $, 
$ \vect{c} = G \vect{c}_0 $, 
$ \vect{d} = G \vect{d}_0 $
with possible repetitions into (\ref{I1})-(\ref{I7}) and 
by making their linear combinations as
\begin{eqnarray}
	{\mathscr F} (G)
	&=&	
	  c_1 I_1 ( G \vect{a}_0, G \vect{a}_0 ) 
	+ c_2 I_1 ( G \vect{a}_0, G \vect{b}_0 ) 
	+ c_3 I_1 ( G \vect{a}_0, G \vect{c}_0 ) 
	+ c_4 I_1 ( G \vect{a}_0, G \vect{d}_0 ) 
	\nonumber \\ &&
	+ c_5 I_1 ( G \vect{b}_0, G \vect{b}_0 ) 
	+ c_6 I_1 ( G \vect{b}_0, G \vect{c}_0 ) 
	+ c_7 I_1 ( G \vect{b}_0, G \vect{d}_0 ) 
	\nonumber \\ &&
	+ c_8 I_1 ( G \vect{c}_0, G \vect{c}_0 ) 
	+ c_9 I_1 ( G \vect{c}_0, G \vect{d}_0 ) 
	+ c_{10} I_1 ( G \vect{d}_0, G \vect{d}_0 ) 
	\nonumber \\ &&
	+ c_{11} I_2 ( G \vect{a}_0, G \vect{a}_0, G \vect{a}_0 ) 
	+ c_{12} I_2 ( G \vect{a}_0, G \vect{a}_0, G \vect{b}_0 ) 
	+ c_{13} I_2 ( G \vect{a}_0, G \vect{a}_0, G \vect{c}_0 ) 
	\nonumber \\ &&
	+ c_{14} I_2 ( G \vect{a}_0, G \vect{a}_0, G \vect{d}_0 ) 
	+ c_{15} I_2 ( G \vect{a}_0, G \vect{b}_0, G \vect{b}_0 ) 
	+ \cdots.
\end{eqnarray}
The coefficients $ c_1, c_2, \cdots $ may depend on temperature or density
of the liquid crystal.
The values of the order parameters, $ G_{abij} $ or $ \hat{G}_{\mu \nu} $,
are determined 
as the solution of the minimization problem of the free energy $ {\mathscr F} (G) $.
However, it can happen that the solution $ G_{abij} $ take physically unrealizable values.
The degrees of order $ \sigma_\alpha $, which were define at (\ref{singular}),
must take their values in the range $ 0 \le \sigma_\alpha \le 1 $
to be physically realizable.
If $ \sigma_\alpha $ is larger than unity, 
the values of the geometric order parameters determined 
by the Landau-de Gennes free energy model 
should be regarded as an unphysical wrong solution.

When an external electric field or magnetic field is applied,
the rotational invariance is broken and hence
the free energy can have extra terms.
If the molecule has an dielectric susceptibility tensor 
$ \vect{s} = (s_{ij}) $
and if an electric field $ \vect{E} $ is applied,
the free energy has an additional term
\begin{equation}
	{\mathscr F}_s (G)
	= \vect{E} \cdot G \vect{s} \cdot \vect{E} 
	= \sum_{a,b,i,j=1}^3 E_a \, E_b \, G_{abij} \, s_{ij}.
\end{equation}
On the other hand,
if the molecule has an electric quadrupole moment 
$ \vect{q} = ( q_{ij} ) $
and if an inhomogeneous electric field $ \vect{E} ( \vect{x} ) $ is applied,
the free energy gets an additional term
\begin{equation}
	{\mathscr F}_q (G)
	= \tr ( \nabla \vect{E} \cdot G \vect{q} )
	= \sum_{a,b,i,j=1}^3 
	\frac{\partial E_a}{\partial x_b} \, G_{abij} \, q_{ij}.
\end{equation}
In most of our discussions we are treating only nematic phases.
Here we briefly discuss other phases 
which are not translationally invariant.
In smectic or cholesteric phases,
the tensorial quantity $ \vect{t} ( \vect{x} ) = G( \vect{x} ) \vect{t}_0 $ 
can depend on the space coordinate $ \vect{x} = (x_1, x_2, x_3) $
and the free energy of a continuum model
has an extra term which is expressed as a spacial integral
\begin{eqnarray}
	{\mathscr F}_k [G] 
	&=&
	\int \sum_{a,b,c=1}^3 \bigg(
	k_1 \, \frac{\partial t_{bc}}{\partial x_a} \, \frac{\partial t_{bc}}{\partial x_a} +
	k_2 \, \frac{\partial t_{bc}}{\partial x_a} \, \frac{\partial t_{ac}}{\partial x_b} 
	\bigg) d^3 x
\nonumber \\
	&=&
	\int \sum_{a,b,c,i,j,k,l=1}^3 \bigg(
	k_1 \frac{\partial G_{bcij}}{\partial x_a} \, \frac{\partial G_{bckl}}{\partial x_a}
	+
	k_2 \frac{\partial G_{bcij}}{\partial x_a} \, \frac{\partial G_{ackl}}{\partial x_b} 
	\bigg) 
	t_{0,ij} \, t_{0,kl} \, d^3 x.
	\label{continuum}
\end{eqnarray}
Then the free energy becomes a functional of $ G_{abij} ( \vect{x} ) $.

Let us apply our general scheme to a system which consists of $ D_{2h} $-symmetric molecules.
The point group $ D_{2h} $ is generated by the set of transformations (\ref{D2h}).
Any microscopic quantity $ \vect{a}_0 $ of a $ D_{2h} $-symmetric molecule 
should satisfy
$ R \vect{a}_0 R^T = \vect{a}_0 $ for $ R \in D_{2h} $.
There are only two independent quantities satisfying this condition,
\begin{equation}
	\vect{a}_0 =
	\frac{1}{\sqrt{6}}
	\begin{pmatrix}
	-1& 0 & 0 \\
	0 &-1 & 0 \\
	0 & 0 & 2
	\end{pmatrix},
	\qquad
	\vect{b}_0 =
	\sqrt{\frac{3}{2}}
	\begin{pmatrix}
	1 & 0 & 0 \\
	0 &-1 & 0 \\
	0 & 0 & 0
	\end{pmatrix}.
\end{equation}
The quantity $ \vect{a}_0 $ is assigned a coordinate
$ ( \varepsilon_1, \varepsilon_2 ) = ( 1, 0 ) $
in the anisotropy diagram.
According to Eq.~(\ref{D2h geo para}),
the coordinate of the corresponding macroscopic quantity 
$ \vect{a} = G \vect{a}_0 $ is
$ ( \tilde{\varepsilon}_1, \tilde{\varepsilon}_2 ) = ( S, \frac{1}{\sqrt{3}} P ) $.
Similarly,
the microscopic quantity $ \vect{b}_0 $ is
$ ( \varepsilon_1, \varepsilon_2 ) = ( 0, \sqrt{3} ) $
and the macroscopic quantity $ \vect{b} = G \vect{b}_0 $ is
$ ( \tilde{\varepsilon}_1, \tilde{\varepsilon}_2 ) = ( D, \frac{1}{\sqrt{3}} C ) $.
They can also be expressed as
\begin{equation}
	\vect{a} =
	\frac{1}{\sqrt{6}}
	\begin{pmatrix}
	-S+P & 0 & 0 \\
	0 & -S-P & 0 \\
	0 & 0 & 2S
	\end{pmatrix},
	\qquad
	\vect{b} =
	\frac{1}{\sqrt{6}}
	\begin{pmatrix}
	-D+C & 0 & 0 \\
	0 &-D-C & 0 \\
	0 & 0 & 2D
	\end{pmatrix}.
\end{equation}
Invariant polynomials formed with these tensors are
\begin{eqnarray}
&&	\tr ( \vect{a} \vect{a} ) = 
	S^2 + \frac{1}{3} P^2,
	\\
&&	\tr ( \vect{a} \vect{b} ) = 
	SD + \frac{1}{3} PC,
	\\
&&	\tr ( \vect{b} \vect{b} ) = 
	D^2 + \frac{1}{3} C^2,
	\\
&&	\tr ( \vect{a} \vect{a} \vect{a} ) = 
	\frac{1}{\sqrt{6}} ( S^3 - S P^2 ),
	\\
&&	\tr ( \vect{a} \vect{a} \vect{b} ) = 
	\frac{1}{3 \sqrt{6}} ( 3 S^2 D - P^2 D - 2 S P C ),
	\\
&&	\tr ( \vect{a} \vect{b} \vect{b} ) = 
	\frac{1}{3 \sqrt{6}} ( 3 S D^2 - S C^2 - 2 P D C ),
	\\
&&	\tr ( \vect{b} \vect{b} \vect{b} ) = 
	\frac{1}{\sqrt{6}} ( D^3 - D C^2 ).
\end{eqnarray}
The relation (\ref{dependence}) implies
\begin{eqnarray}
&&	\tr ( \vect{a} \vect{a} \vect{a} \vect{a} ) 
	- \frac{1}{2} \tr ( \vect{a} \vect{a} ) \, \tr ( \vect{a} \vect{a} ) = 0,
	\\
&&	\tr ( \vect{a} \vect{a} \vect{a} \vect{b} ) - 
	\frac{1}{2} \tr ( \vect{a} \vect{a} ) \, \tr ( \vect{a} \vect{b} ) = 0,
	\\
&&	2 \tr ( \vect{a} \vect{a} \vect{b} \vect{b} ) 
	+ \tr ( \vect{a} \vect{b} \vect{a} \vect{b} ) 
	- \tr ( \vect{a} \vect{b} ) \, \tr ( \vect{a} \vect{b} )
	- \frac{1}{2} \tr ( \vect{a} \vect{a} ) \, \tr ( \vect{b} \vect{b} ) = 0.
\end{eqnarray}
Furthermore, since product of diagonal tensors $ \vect{a}, \vect{b} $ is commutative as
$ \vect{b} \vect{a} = \vect{a} \vect{b} $,
it holds that
\begin{equation}
	\tr ( \vect{a} \vect{a} \vect{b} \vect{b} ) 
	= \tr ( \vect{a} \vect{b} \vect{a} \vect{b} ).
\end{equation}
Hence there are only six independent invariants of the fourth order
\begin{eqnarray}
&&	\tr ( \vect{a} \vect{a} ) \tr ( \vect{a} \vect{a} ),
	\quad
	\tr ( \vect{a} \vect{a} ) \tr ( \vect{a} \vect{b} ),
	\quad
	\tr ( \vect{a} \vect{a} ) \tr ( \vect{b} \vect{b} ),
	\nonumber \\
&&	\tr ( \vect{a} \vect{b} ) \tr ( \vect{a} \vect{b} ),
	\quad
	\tr ( \vect{a} \vect{b} ) \tr ( \vect{b} \vect{b} ),
	\quad
	\tr ( \vect{b} \vect{b} ) \tr ( \vect{b} \vect{b} ).
\end{eqnarray}
The Landau-de Gennes free energy of $ D_{2h} $-symmetric molecules is constructed
as a linear combination of these invariants,
\begin{eqnarray}
	{\mathscr F} (S,P,D,C)
&=&	c_1 \tr ( \vect{a} \vect{a} ) +
	c_2 \tr ( \vect{a} \vect{b} ) +
	c_3 \tr ( \vect{b} \vect{b} ) 
	\nonumber \\ && + 
	c_4 \tr ( \vect{a} \vect{a} \vect{a} ) + 
	c_5 \tr ( \vect{a} \vect{a} \vect{b} ) + 
	c_6 \tr ( \vect{a} \vect{b} \vect{b} ) + 
	c_7 \tr ( \vect{b} \vect{b} \vect{b} ) 
	\nonumber \\ && + 
	c_8 \tr ( \vect{a} \vect{a} ) \tr ( \vect{a} \vect{a} ) +
	c_9 \tr ( \vect{a} \vect{a} ) \tr ( \vect{a} \vect{b} ) +
	c_{10} \tr ( \vect{a} \vect{a} ) \tr ( \vect{b} \vect{b} ) 
	\nonumber \\ && + 
	c_{11} \tr ( \vect{a} \vect{b} ) \tr ( \vect{a} \vect{b} ) +
	c_{12} \tr ( \vect{a} \vect{b} ) \tr ( \vect{b} \vect{b} ) +
	c_{13} \tr ( \vect{b} \vect{b} ) \tr ( \vect{b} \vect{b} )
	\nonumber \\
&=&	
	  c_1 ( S^2 + \frac{1}{3} P^2 ) 
	+ c_2 ( SD + \frac{1}{3} PC )
	+ c_3 ( D^2 + \frac{1}{3} C^2 )
	\nonumber \\ && 
	+ \frac{c_4}{\sqrt{6}} ( S^3 - S P^2 )
	+ \frac{c_5}{3 \sqrt{6}} ( 3 S^2 D - P^2 D - 2 S P C )
	\nonumber \\ && 
	+ \frac{c_6}{3 \sqrt{6}} ( 3 S D^2 - S C^2 - 2 P D C )
	+ \frac{c_7}{\sqrt{6}} ( D^3 - D C^2 )
	\nonumber \\ && 
	+ c_8
	( S^2 + \frac{1}{3} P^2 )^2
	+ c_9
	( S^2 + \frac{1}{3} P^2 )
	( SD + \frac{1}{3} PC )
	+ c_{10}
	( S^2 + \frac{1}{3} P^2 )
	( D^2 + \frac{1}{3} C^2 )
	\nonumber \\ && 
	+ c_{11}
	( SD + \frac{1}{3} PC )^2
	+ c_{12}
	( SD + \frac{1}{3} PC )
	( D^2 + \frac{1}{3} C^2 )
	+ c_{13}
	( D^2 + \frac{1}{3} C^2 )^2.
	\label{LG free energy}
\end{eqnarray}
up to the fourth order.
This contains more terms than the free energy formulated by
Allender {\it et al.}~\cite{Allender1984, Allender1985}
even if only terms lower than fifth order are compared.
Calculation of higher order terms is cumbersome but feasible.
The values of the order parameters $ S,P,D,C $ are determined
as a solution of the minimization problem of the free energy $ {\mathscr F} (S,P,D,C) $.
It should be checked whether these values are physically realizable or not.
We can calculate singular values $ \sigma_\alpha $ $ ( \alpha = 1,2 ) $
of the reduced ordering matrix $ \hat{G} $, which was defined at (\ref{D2h geo para}),
\begin{equation}
	\hat{G}
	=
	\begin{pmatrix}
	\hat{G}_{55} &
	\hat{G}_{54} \\
	\hat{G}_{45} &
	\hat{G}_{44} 
	\end{pmatrix}
	=
	\begin{pmatrix}
	S                    & \frac{1}{\sqrt{3}} D \\
	\frac{1}{\sqrt{3}} P & \frac{1}{3} C
	\end{pmatrix}.
	\label{D2h geo para matrix}
\end{equation}
The singular values $ \sigma_\alpha $ should be in the range
$ | \sigma_\alpha | \le 1 $.
If they are not in this range, the values of $ S,P,D,C $ are physically unrealizable.
In such a case, the coefficients $ c_1, c_2, \cdots $ in the polynomial $ {\mathscr F} $
should be re-adjusted.

It should be noted that the solution of 
the minimization problem of $ {\mathscr F} (S,P,D,C) $ is not unique.
As a remnant of the rotational symmetry of the free energy,
the solutions have the permutation symmetry $ {\frak S}_3 $, which is generated
by the rotation matrices $ \{ K_1, K_2, K_3 \} $ given at (\ref{Ks}),
or by the transformations of the anisotropy diagram
$ \{ \phi_1, \phi_2, \phi_3 \} $ given at (\ref{phi_1})-(\ref{phi_3}).
If $ (S,P,D,C) $ is a solution of the minimization problem,
$ (S',P',D',C') $ which is defined by
\begin{eqnarray}
	\begin{pmatrix}
	S'                    & \frac{1}{\sqrt{3}} D' \\
	\frac{1}{\sqrt{3}} P' & \frac{1}{3} C'
	\end{pmatrix}
	=
	\phi_1 \hat{G}
&=&
	\frac{1}{2}
	\begin{pmatrix}
	-1 & - \sqrt{3} \\
	- \sqrt{3} & 1
	\end{pmatrix}
	\begin{pmatrix}
	S                    & \frac{1}{\sqrt{3}} D \\
	\frac{1}{\sqrt{3}} P & \frac{1}{3} C
	\end{pmatrix}
\nonumber \\
&=&
	\frac{1}{2}
	\begin{pmatrix}
	-S-P                    & - \frac{1}{\sqrt{3}} D - \frac{1}{\sqrt{3}} C \\
	- \sqrt{3} S + \frac{1}{\sqrt{3}} P & -D + \frac{1}{3} C
	\end{pmatrix}
\end{eqnarray}
is also a solution.
In this way we obtain a complete set of equivalent solutions,
$ \{ \hat{G}, $ 
$ \phi_1 \hat{G}, $ 
$ \phi_2 \hat{G}, $ 
$ \phi_3 \hat{G}, $ 
$ \phi_2 \phi_1 \hat{G}, $ 
$ \phi_1 \phi_2 \hat{G} \} $,
which are written as
\begin{eqnarray}
	(S', P', D', C' )
&=&
	(S, P, D, C ),
	\nonumber \\ &&
	\frac{1}{2} (-S-P, -3S+P, -D-C, -3D+C ),
	\nonumber \\ &&
	\frac{1}{2} (-S+P, 3S+P, -D+C, 3D+C ),
	\nonumber \\ &&
	(S, -P, D, -C ),
	\nonumber \\ &&
	\frac{1}{2} (-S+P, -3S-P, -D+C, -3D-C ),
	\nonumber \\ &&
	\frac{1}{2} (-S-P, 3S-P, -D-C, 3D-C ).
\end{eqnarray}

\section{Conclusion}\label{sect9}
In the introduction of this paper
we pointed out that the conventional method 
using tensorial order parameters (\ref{A}), (\ref{B}), (\ref{C})
for characterizing biaxial nematic phases becomes ambiguous
when it is applied to a system of asymmetric molecules.
Since the conventional tensorial order parameters depend on 
the choice of a reference frame fixed on the molecule
and an asymmetric molecule does not have preferable axes,
the order parameters are not defined uniquely.
What is worse, an asymmetric molecule may possess
various tensorial physical quantities which do not have common principal axes.
Although the ordering matrix~(\ref{de Gennes}),
which was originally introduced by de Gennes,
is applicable to a molecule which has an arbitrary shape,
the interpretation of the ordering matrix is difficult.
Thus, we aimed to invent useful tools 
for describing and for analyzing geometric structures of biaxial nematics.

Here we summarize the main results of this paper.
Around Eq.~(\ref{micro-macro}) we argued that the ordering matrix 
is to be understood as the geometric order parameter 
$ G_{abij} = \bra Q_{abij} \ket $
which relates
the microscopic quantity $ \vect{t} $ intrinsic in a molecule
to the macroscopic quantity $ \bra \tilde{\vect{t}} \ket = G \vect{t} $
observed in a bulk system.
The geometric order parameter was analyzed by the singular value decomposition.
At Eq.~(\ref{singular}) it was shown that
the microscopic singular tensor $ \vect{t}_\alpha $
manifests itself as the macroscopic singular tensor $ \vect{u}_\alpha $
in the nematic phase with the strength of realization $ \sigma_\alpha $.

Any tensorial quantity is mapped in the anisotropy diagram.
It should be noted that
six or three equivalent points in the anisotropy diagram
correspond to one tensor.
As indices for evaluating anisotropies of tensorial quantities,
we introduced the degree of anisotropy $ \alpha $ at Eq.~(\ref{alpha})
and the degree of biaxiality $ \beta $ at Eq.~(\ref{beta}).
The index $ \alpha $ is the radius in the anisotropy diagram
and $ \beta $ is the angle measured 
from the uniaxial line in the anisotropy diagram.
By proving theorems 1, 2 and 3
we showed the micro-macro relation, which tells that
the point representing the macroscopic tensor always locates
in the polygon in the diagram 
whose vertices are points representing the microscopic tensor.

In Sect.~\ref{sect7} we applied our method to a system which consists of
$ D_{2h} $-symmetric molecules.
All the tensorial quantities of a $ D_{2h} $-symmetric molecule 
have common principal axes and hence they are simultaneously diagonalizable.
The geometric order parameter also becomes diagonal,
namely, only the components $ G_{aaii} $ with $ a,i =1,2,3 $ can be nonzero.
Hence it has only four independent components as shown in (\ref{solved}).
In this case the micro-macro quantities are related as (\ref{D2h geo para}).

In Sect.~\ref{sect8} we explained the general prescription 
to formulate the Landau-de Gennes free energy.
By this prescription we can construct the free energy
which contains all the symmetry-admissible terms but contains no redundant terms.
We wrote down the concrete Landau-de Gennes free energy (\ref{LG free energy})
for the $ D_{2h} $-symmetric molecule system.

We would like to emphasize that we made 
the implication of de Gennes's ordering matrix clear 
by interpreting it as the geometric order parameter which transforms
a microscopic quantity to a macroscopic quantity.
The anisotropy diagram and the anisotropy indices which we introduced
are systematic tool and help us understand the properties of biaxial nematics.
It also should be emphasized that our method has no ambiguity.
As an example to show the uniqueness of our procedure,
we formulated the Landau-de Gennes free energy 
in the most general form without redundancy.

Here we mention some remaining problems.
We should analyze the phase structure using the Landau-de Gennes free energy.
This problem will be discussed in the next work.
Our method can be applied also for analysis of molecular dynamics simulation of nematics.
It can be generalized for treating a system 
which is a mixture of rod-shaped molecules and disk-shaped molecules.
It can be generalized for treating flexible molecules
although in our discussion molecules were assumed to be rigid.
It is also interesting to study smectic phases using the continuum model
which was briefly discussed at (\ref{continuum}).
Kimura~\cite{Kimura1974} studied a system of rod-shaped molecules 
which interact each other
via both the short-range exclusion force and the long-range dispersion force.
For an asymmetric molecule
the principal axes of its electric quadrupole tensor
may not coincide with 
the geometric axes which characterize the rigid-body repulsive force.
It seems interesting to study a system which consists of such asymmetric molecules.
In this paper we analyzed only the second-rank tensors.
It is possible to extend our argument to include higher-rank tensorial quantities
although necessary for such an extension is not obvious 
in the context of liquid crystal physics.

\section*{Acknowledgements}
Tanimura and Koda would like to thank Dr. Shohei Naemura
for stimulating and insightful discussions with him.
This work is partly supported by the Global COE program 
``Informatics Center for the Development of Knowledge Society Infrastructure''
of Kyoto University, 
and also by the Grant-in-Aid for Scientific Research on Priority Area ``Soft Matter Physics'' 
of the Ministry of Education, Culture, Sports, Science and Technology of Japan.

\appendix
\section{Proof of theorem 4}
Here we prove that Eqs.~(\ref{I1})-(\ref{I8}) are a complete list
of invariant polynomials up to the fourth order.
We also prove Eq.~(\ref{dependence}),
\begin{equation}
	I_3 + I_4 + I_5 - \frac{1}{2} ( I_6 + I_7 + I_8 ) = 0.
	\label{appendix}
\end{equation}
First, we count linearly independent invariant polynomials formed by
products of traceless symmetric tensors.
The set of the whole traceless symmetric tensors becomes 
a five-dimensional irreducible representation space of the rotation group.
The symbol $ {\mathbf 5} $ or $ {\mathbf 3} $ denotes
a five-dimensional or three-dimensional irreducible representation space, respectively.
The one-dimensional representation space $ {\mathbf 1} $ is a set of quantities
which are invariant under the action of the rotation group.
In other words, $ {\mathbf 1} $ is a set of scalars.
According to the Clebsch-Gordan law~\cite{Sakurai1985},
the tensor prodct space $ {\mathbf 5} \otimes {\mathbf 5} $ is decomposed as
\begin{equation}
	{\mathbf 5} \otimes {\mathbf 5} = 
	{\mathbf 1} \oplus {\mathbf 3} \oplus {\mathbf 5} \oplus {\mathbf 7} \oplus {\mathbf 9}.
\end{equation}
In the decomposition the one-dimensional representation $ {\mathbf 1} $ appears once,
which corresponds to
$ I_1 = \tr ( \vect{a} \vect{b} ) $ of Eq.~(\ref{I1}).
Similarly, the three-fold tensor product 
$ {\mathbf 5} \otimes {\mathbf 5} \otimes {\mathbf 5} $ is decomposed as
\begin{eqnarray}
	{\mathbf 5} \otimes {\mathbf 5} \otimes {\mathbf 5} 
	& = &
	{\mathbf 5} \oplus 
	\nonumber \\ &&
	{\mathbf 3} \oplus {\mathbf 5} \oplus {\mathbf 7} \oplus 
	\nonumber \\ &&
	{\mathbf 1} \oplus {\mathbf 3} \oplus {\mathbf 5} \oplus {\mathbf 7} \oplus {\mathbf 9} \oplus
	\nonumber \\ &&
	{\mathbf 3} \oplus {\mathbf 5} \oplus {\mathbf 7} \oplus {\mathbf 9} \oplus {\mathbf{11}} \oplus 
	\nonumber \\ &&
	{\mathbf 5} \oplus {\mathbf 7} \oplus {\mathbf 9} \oplus {\mathbf{11}} \oplus {\mathbf{13}}.
\end{eqnarray}
In this decomposition $ {\mathbf 1} $ appears only once,
which corresponds to
$ I_2 = \tr ( \vect{a} \vect{b} \vect{c} ) $ of Eq.~(\ref{I2}).
The calculation of the four-fold tensor product yields
\begin{eqnarray}
	{\mathbf 5} \otimes {\mathbf 5} \otimes {\mathbf 5} \otimes {\mathbf 5} 
	& = &
	{\mathbf 1} \oplus {\mathbf 3} \oplus {\mathbf 5} \oplus {\mathbf 7} \oplus {\mathbf 9} \oplus
	\nonumber \\ &&
	{\mathbf 3} \oplus {\mathbf 5} \oplus {\mathbf 7} \oplus 
	\nonumber \\ &&
	{\mathbf 1} \oplus {\mathbf 3} \oplus {\mathbf 5} \oplus {\mathbf 7} \oplus {\mathbf 9} \oplus
	\nonumber \\ &&
	{\mathbf 3} \oplus {\mathbf 5} \oplus {\mathbf 7} \oplus {\mathbf 9} \oplus {\mathbf{11}} \oplus 
	\nonumber \\ &&
	{\mathbf 5} \oplus 
	\nonumber \\ &&
	{\mathbf 3} \oplus {\mathbf 5} \oplus {\mathbf 7} \oplus 
	\nonumber \\ &&
	{\mathbf 1} \oplus {\mathbf 3} \oplus {\mathbf 5} \oplus {\mathbf 7} \oplus {\mathbf 9} \oplus
	\nonumber \\ &&
	{\mathbf 3} \oplus {\mathbf 5} \oplus {\mathbf 7} \oplus {\mathbf 9} \oplus {\mathbf{11}} \oplus 
	\nonumber \\ &&
	{\mathbf 5} \oplus {\mathbf 7} \oplus {\mathbf 9} \oplus {\mathbf{11}} \oplus {\mathbf{13}} \oplus
	\nonumber \\ &&
	{\mathbf 3} \oplus {\mathbf 5} \oplus {\mathbf 7} \oplus 
	\nonumber \\ &&
	{\mathbf 1} \oplus {\mathbf 3} \oplus {\mathbf 5} \oplus {\mathbf 7} \oplus {\mathbf 9} \oplus
	\nonumber \\ &&
	{\mathbf 3} \oplus {\mathbf 5} \oplus {\mathbf 7} \oplus {\mathbf 9} \oplus {\mathbf{11}} \oplus 
	\nonumber \\ &&
	{\mathbf 5} \oplus {\mathbf 7} \oplus {\mathbf 9} \oplus {\mathbf{11}} \oplus {\mathbf{13}} \oplus
	\nonumber \\ &&
	{\mathbf 7} \oplus {\mathbf 9} \oplus {\mathbf{11}} \oplus {\mathbf{13}} \oplus {\mathbf{15}} \oplus
	\nonumber \\ &&
	{\mathbf 1} \oplus {\mathbf 3} \oplus {\mathbf 5} \oplus {\mathbf 7} \oplus {\mathbf 9} \oplus
	\nonumber \\ &&
	{\mathbf 3} \oplus {\mathbf 5} \oplus {\mathbf 7} \oplus {\mathbf 9} \oplus {\mathbf{11}} \oplus 
	\nonumber \\ &&
	{\mathbf 5} \oplus {\mathbf 7} \oplus {\mathbf 9} \oplus {\mathbf{11}} \oplus {\mathbf{13}} \oplus
	\nonumber \\ &&
	{\mathbf 7} \oplus {\mathbf 9} \oplus {\mathbf{11}} \oplus {\mathbf{13}} \oplus {\mathbf{15}} \oplus
	\nonumber \\ &&
	{\mathbf 9} \oplus {\mathbf{11}} \oplus {\mathbf{13}} \oplus {\mathbf{15}} \oplus {\mathbf{17}}.
\end{eqnarray}
In this decomposition $ {\mathbf 1} $ appears five times.
Thus there must be five linearly independent invariants of the fourth order
and there are no more than five.
The six quantities listed in Eqs.~(\ref{I3})-(\ref{I8}) are invariants of the fourth order.
By construction it is obvious that there are no more independent polynomials 
of the fourth order.
Hence the six quantities must have one nontrivial relation.
{}From their symmetry, we can guess a relation of the form
\begin{equation}
	I_3 + I_4 + I_5 + c ( I_6 + I_7 + I_8 ) = 0.
	\label{guess}
\end{equation}
If we substitute
\begin{equation}
	\vect{a} = \vect{b} = \vect{c} = \vect{d} =
	\begin{pmatrix}
	-1& 0 & 0 \\
	0 &-1 & 0 \\
	0 & 0 & 2
	\end{pmatrix},
\end{equation}
we get 
$ I_3 = I_4 = I_5 = 18 $ and $ I_6 = I_7 = I_8 = 36 $.
Hence the coefficient $ c $ in (\ref{guess}) must be $ c = - \frac{1}{2} $.
This proves (\ref{dependence}).
If another evidence is requested, we may substitute
\begin{equation}
	\vect{a} = \vect{b} = \vect{c} = \vect{d} =
	\begin{pmatrix}
	1 & 0 & 0 \\
	0 &-1 & 0 \\
	0 & 0 & 0
	\end{pmatrix}.
\end{equation}
Then we get 
$ I_3 = I_4 = I_5 = 2 $ and $ I_6 = I_7 = I_8 = 4 $.
This confirms that $ c = - \frac{1}{2} $.
The reader may calculate other cases to confirm (\ref{appendix}).

\end{document}